\documentclass[10pt,conference]{IEEEtran}
\usepackage{cite}
\usepackage{amsmath,amssymb,amsfonts}

\usepackage{graphicx}
\usepackage{textcomp}
\usepackage{xcolor}
\usepackage[hyphens]{url}
\usepackage{fancyhdr}
\usepackage{hyperref}

\usepackage{algorithm}
\usepackage{algpseudocode}
\usepackage{comment}
\newcommand{\hpcayear}{2027}

\newcommand{\red}[1]{\textcolor{red}{#1}}

\newcommand{\ignore}[1]{}

\newcommand{\arch}{PRIM}

\title{\huge {\arch: Energy- and Endurance-Aware Compiler Optimization for Digital Compute in Non-volatile Memory}}
\def\hpcacameraready{} % Uncomment to build camera-ready version

\newcommand\hpcaauthors{Patrick Miller and Ulya R. Karpuzcu}
\newcommand\hpcaaffiliation{University of Minnesota, Twin Cities, USA}
\author{
  \ifdefined\hpcacameraready
    \IEEEauthorblockN{\hpcaauthors{}}
      \IEEEauthorblockA{
        \hpcaaffiliation{} \\
      }
  \else

  \fi 
}

\begin{document}
\maketitle

%Enables the camera ready header and footer
\ifdefined\hpcacameraready 
  \pagestyle{empty}
\else
  \thispagestyle{plain}
  \pagestyle{plain}
\fi

\newcommand{\hpcaheight}{0mm}
\ifdefined\eaopen
\renewcommand{\hpcaheight}{12mm}
\fi

%%%%%%%%%%%%%%%%%%%%%%%%%%%%%%%%%%%%%%%%
%%%%%%%% -- PAPER CONTENT STARTS -- %%%%%%%%%

\begin{abstract}
Compute-in-Memory (CiM) is an alternative computing paradigm that circumvents the memory bandwidth bottleneck. Variants utilizing non-volatile memory cells (NVCiM) to perform Boolean operations directly within memory arrays offer significant density and energy-efficiency advantages. Fully realizing this potential, however, requires NVCiM-specific compiler optimizations. In conventional digital NVCiM systems, an output cell must be written---or preset—to an operation-specific value before any Boolean operation can occur. Preset-induced extra writes can degrade energy efficiency and cell endurance, often to the point of negating CiM's efficiency gains.
In this paper, we introduce PRIM (Preset Reduction In Memory), a compiler optimization framework tailored for digital NVCiM that effectively minimizes preset overhead without compromising computational accuracy or architectural complexity. Unlike existing optimization approaches, PRIM is not restricted to specific types of Boolean operations. Experimental results show that PRIM can reduce execution time and energy consumption by up to 50\% and 44\%, respectively, while boosting cell endurance by up to 50\%.
 %Compute in Memory (CiM) is an alternative computing paradigm that circumvents the performance bottleneck of limited memory bandwidth. Variants using non-volatile memory cells (NVCiM) to directly perform Boolean operations (gates or functions) in memory arrays provide significant density and energy efficiency benefits.Realizing this full potential relies critically on NVCiM-specific performance optimizations. In conventional digital NVCiM systems, before any Boolean operation takes place, the cell serving as the output must be written---{\em preset to}---an operation-specific value. Preset induced writes are extra writes that reduce energy efficiency and worsen cell endurance---to the extent to wipe out energy efficiency gains from CiM. In this paper, we introduce a digital NVCiM-specific compiler optimization framework, \arch\ (Preset Reduction In Memory), which can effectively minimize the preset overhead without compromising computational accuracy or architectural complexity.We demonstrate that \arch\ can result in up-to {50}\% ({44}\%) reduction in execution time (energy consumption) along with endurance improvement by up-to {50}\%. Contrary to similar optimization approaches, \arch\ is unique in that it is not restricted to a specific type of Boolean operation.
\end{abstract}

\section{{Introduction}}
%\blue{Better name for this kind of optimization? Been calling it preset optimization, but there may be a more distinctive or compact name. There are hardware designs that use sequential logic to prevent combinational logic blocks from growing too large or to save power. FPGAs can also do this. I have not found what terminology is used for this.}
%\blue{Note: need to evaluate font directions from ISCA main.txt. How to determine font size? Shrink space below figs? Extending data to perhaps replace the variable STT resistance section.}
% Nothing needed, commenting this.
%\blue{Question: when referring to figures, the paper is using Figure and Fig., what is best for consistency?}\red{Fig. if we are short in space -- doesn't seem to be the case now, Figure otherwise}

Performance scaling in emerging applications of practical importance
%many big data applications 
such as genomics \cite{bigDataGenomics} or artificial intelligence \cite{aiMemWall} is limited by the memory bandwidth.
%available to the system. 
One promising computational paradigm is the integration of processing capability into or near memory, known as Compute in Memory (CiM). 
%A range of CiM architectures have been proposed to accelerate these memory-intensive workloads. 
%\cite{nejatollahi2020cryptopim, cilasun2020crafft, imani2019floatpim, khalifa2021filtpim}.
%One subset of the design space involves 
In the rich design space of CiM, 
architectures utilizing non-volatile memory-based CiM (NVCiM) stand out \cite{nejatollahi2020cryptopim, cilasun2020crafft, imani2019floatpim, khalifa2021filtpim}, where the storage arrays of a persistent memory technology are modified to perform digital computation by directly using the memory cells.
%themselves. 
The non-volatile memory provides density and energy benefits, and utilizing memory cells as compute elements enables the tightest possible coupling between computing and memory, which in turn maximizes bandwidth and minimizes data movement. 
%and avoiding reliance on 
At the same time, these CiM systems do not need 
complex analog or special-purpose circuitry at the array periphery.

In order to perform computation, NVCiM systems equip memory arrays with a few extra control lines
%the memory arrays are modified 
to connect multiple memory cells to form a logic gate -- typically one or more cells serve as the inputs; another cell, as the output. All such cells reside either in the same row or in the same column.  
%One or more cells are selected as inputs, and they are connected to a cell selected as the output. 
Before the computation starts, the output cell ${Out}$ is {\em preset}, i.e., a gate-specific, known value $v_{out}$ is written to the output cell. Once the computation is fired using dedicated control lines, the output cell may or may not change state depending on the logic states of the input cells. %, according to the truth table of the targeted gate operation.  
The purpose of presetting the output cell to a known value is to enable deterministic  
%More specifically, 
%This method allows for 
%the output cell's 
${Out}$ state transitions  %to 
matching 
%mapping to 
the truth table of the targeted gate, which is crucial for correctness. 

Typical NVCiM architectures, as such, can support 
universal Boolean gates. Therefore, we can map any digital computation to NVCiM by translating the underlying computational steps into a sequence of Boolean operations.
%such as NAND or NOR, allowing for the implementation of algorithms by synthesizing sequences of Boolean operations to be executed by the CiM. 
At the same time, all rows or columns in the memory array can perform the same operation on different data simultaneously, and multiple arrays can compute in parallel. 
%Further, the many rows or columns in a memory array can provide significant opportunities for parallel execution, by allowing rows of cells to execute the same operation.
%is written 
%set to a predetermined state, and then the mechanism used for overwriting the output cell's state is modulated by the states of the input cells. 
The very same method to implement logic operations applies across different NVM technologies
%to be connected. 
%The specific NVM technology, 
such as Resistive Random Access Memory (ReRAM) or Magnetoresistive  Random Access Memory (MRAM), 
which otherwise rely on different physical phenomena to represent logic states \cite{nejatollahi2020cryptopim, cilasun2020crafft, imani2019floatpim, khalifa2021filtpim}. 
%vary in 
%the physical phenomena they rely upon for state storage, but the method for implementing logic gates is shared. 

%A practical roadblock to harvesting digital NVCiM's full energy efficiency and performance potential is the overhead of preset operations. Each logic operation must start with the preset (write) to the output cell, which incurs a significant time and energy overhead by definition. The overall preset overhead grows with the number of presets necessary to execute an application using NVCiM. 
%
%Worse, non-volatile memory technologies have finite write endurance, which limits the number of writes to each cell during the lifetime of the memory array. Output cells in particular already experience accelerated wear due to logic operations repeatedly overwriting their state. Additional writes induced by presets can only exacerbate this situation. 

A major bottleneck to fully realizing the energy-efficiency and performance potential of digital NVCiM is the overhead imposed by preset operations. Because each logic operation requires presetting (writing) the destination cell, it inherently incurs substantial latency and energy penalties. Consequently, the total preset overhead scales directly with the number of preset operations required by an application. Compounding this problem, non-volatile memory technologies exhibit finite write endurance, which limits the number of writes to each cell during the lifetime of the memory array. Since output cells already undergo accelerated wear from repeated logic updates, additional preset-induced writes can significantly degrade overall lifetime.
 
%Typical NVCiM compilers seek to minimize the total number of logic operations representing application-level computational steps, $N_{op}$, as the end-to-end execution time and energy consumption are directly proportional to $N_{op}$. 
%the number of gate operations that have been synthesized 
%State-of-the-art NVCiM compilers, however, treat each preset as an atomic component of a gate operation, which, as we demonstrate in this paper, leaves significant optimization opportunities untapped. In fact, we observe that numerous presets over long chains of dependent operations can safely be eliminated through operation scheduling or reordering. 

Typical NVCiM compilers focus on minimizing the total number of logic operations, $N_{op}$, because end-to-end execution time and energy consumption scale directly with $N_{op}$. However, state-of-the-art compilers treat each preset as an atomic component tightly bound to a gate operation—a design choice that leaves significant optimization opportunities untapped. Crucially, we observe that numerous presets across long chains of dependent operations can be safely eliminated through strategic operation scheduling and reordering.

%Simply eliminating a given preset, i.e., the dedicated write to an output cell $cell_{out}$, would not be safe for correctness. In the end, to implement a logic operation we need to have $cell_{out}$ set to a operation-specific logic value $v_{out}$ before we fire computation. In the absence of the dedicated preset, we can guarantee correctness only if $cell_{out}=v_{out}$ applies before computation starts. The core idea behind \arch\ is facilitating this, through operation reordering/scheduling, by letting predecessor operations with an output value of $v_{out}$ in program order to set $cell_{out}$, as opposed to a dedicated preset (write) to $cell_{out}$.  
%
%Safe elimination of presets directly improves performance, energy efficiency and endurance. At the same time, without frequent preset intervention, as we will detail in the following, maintaining  ``sticky'' bits, forming n($>3$)-input gates using 2- or 3-input gates as building blocks, or scanning bit vectors become much easier to implement---which represent common useful NVCiM primitives. 
%presets making the maintenance of certain kinds of state such as ``sticky'' bits, synthesizing many-input gates from smaller variants, or scanning bit vectors more difficult.

Naively eliminating a preset---the dedicated write to an output cell (${Out}$)---compromises functional correctness. Executing a logic operation requires ${Out}$ to hold an operation-specific initial value, $v_{out}$, prior to computation. In the absence of an explicit preset, correctness can only be guaranteed if ${Out}=v_{Out}$ already holds before execution begins.
The core principle behind PRIM is to fulfill this requirement through operation reordering and scheduling. Specifically, PRIM identifies operations whose ${Out}$, if shared as the output of subsequent operations, leads to modified behaviors that recreate the desired outcome. This bypasses the need for the later preset writes, and enables further optimization opportunities for eliminating and/or coalescing remaining logic operations.

Safely eliminating presets directly improves performance, energy efficiency, and cell endurance. Furthermore, reducing preset interventions simplifies in-memory implementation of several useful NVCiM primitives which require keeping a history of computation results from consecutive execution steps---such as maintaining sticky bits, constructing n-input logic gates ($n>3$) from 2- or 3-input building blocks, and scanning bit vectors.

%While this method provides some significant potential benefits, there are trade-offs that can impact the practicality of such devices. There is significant pressure to optimize the Boolean sequences implementing various functions, as the execution latency and energy cost of a function is proportional to the number of gate operations that have been synthesized. In addition to this, non-volatile memory technologies have finite write endurance, and output cells in particular experience accelerated wear due to gate operations repeatedly overwriting their state.

%The two-step process of setting the output cell and then performing a gate operation presents an optimization target for algorithms implemented in a NVCiM to reduce the time, power, and endurance loss related to the preset step. Analysis of the Boolean operations needed to implement functions in a CiM can reveal where presets or intermediate gate operations can be skipped, if a chain of dependent instructions without intervening presets leads to the same output.

%This change in methodology leverages the existing functionality of a NVCiM array, allowing for minimal alterations to improve performance, energy, and endurance of various common instruction sequences, and providing significant improvement on some Boolean sequences that expand significantly when implemented on a CiM due to presets making the maintenance of certain kinds of state such as ``sticky'' bits, synthesizing many-input gates from smaller variants, or scanning bit vectors more difficult.

In the following, we cover fundamental
%background 
principles and challenges of NVCiM
%peration and its challenges 
in Section \ref{sec:bkg};
basics of \arch,  in 
%describe the optimization process and discuss versions of common compute blocks (
Section \ref{sec:design};
%, describe the 
experimental setup, in Section \ref{sec:eval_setup}; 
a quantitative characterization of energy efficiency and endurance implications, in
%evaluate the impacts on energy and endurance (
Section \ref{sec:eval}; and related work in 
%, discuss related works (
Section \ref{sec:relwork}.
%), and discuss technology projections (Section \ref{sec:projections}).
%\blue{prune}

%\section{{Background}}
\section{{Digital NVCiM Basics}}
\label{sec:bkg}
\ignore{
\begin{figure}[tbp]
\centerline{\includegraphics[scale=0.6]{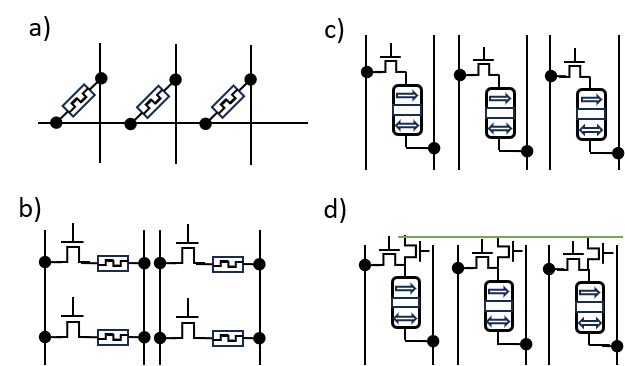}}
\caption{
Non-volatile Array Structures 
a) Crossbar ReRAM layout. b) 1T1R ReRAM array.
c) MRAM layout. d) NV-CiM MRAM layout.
}
%\vspace{-0.3cm}
\label{fig:array_comp}
\end{figure}
}

\begin{figure}[tbp]
\centerline{\includegraphics[scale=0.45]{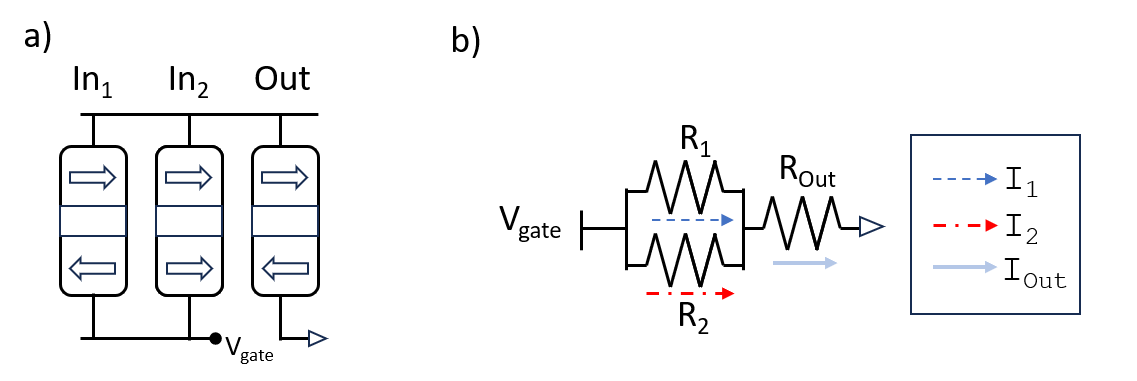}}
\vspace{-.1cm}
\caption{Digital NVCiM:
%MRAM. 
(a) Logic gate formation. 
%Arrows in cells show magnetic orientation of MTJ fixed and free layers. $In_1$ and $In_2$ represent the inputs; $Out$, the total current of the gate controlled by a specific voltage $V_{gate}$. 
(b) Electrical equivalent of a logic gate; a resistive network. 
%$I_1$, $I_2$ and $I_{Out}$ are currents through cells $In_1$, $In_2$ and $Out$, which are the inputs and output, respectively.
}
\vspace{-0.6cm}
\label{fig:cram_basics}
\end{figure}

\noindent{\bf Operation Principle:}
State-of-the-art digital NVCiM architectures come in two flavors: ReRAM-based \cite{magic,imani_digitalpim_2019,simplermagic,xmagic,iloss,scrimp20,felix2018} and MRAM-based \cite{wangsotserial,zabihi2018memory,resch2020mouse,resch2023pimcity}. 
%\red{[would be good to add a few more citations for each case, from diverse groups, please also check the existing ones]}.
%\blue{[added more]}
%\red{thx, can we add some other refs than CRAM under MRAM?} \blue{CRAM seems to be common for MRAM, found one slightly different MRAM design, BSCIM. Many others modify topology significantly. }
%\red{can we add a UCSD design here, first author Imani?}\blue{added as second RRAM citation}
Fig.\ref{fig:cram_basics} covers the basics for MRAM-based 
NVCiM, which, without loss of generality, 
%is modeled on an MRAM device and 
serves as the baseline throughout this paper. The very same principle of gate formation applies to ReRAM-based NVCiM.

Fig.1(a) depicts the MRAM cell configuration for a generic logic operation---with $In_1$ and $In_2$ as inputs, and $Out$ as the output.
Data is stored in Magnetic Tunnel Junctions (MTJs) consisting of a fixed reference layer and a switchable free layer (magnetic orientations indicated by arrows). Binary states are encoded via MTJ resistance: a low-resistance parallel state ($R_{P}$) represents logic `0'; a high-resistance antiparallel state ($R_{AP}$), logic `1'. Switching the free layer requires a current exceeding the critical threshold ($I_{crit}$), with current direction dictating the final state.

%The electrical equivalent of the modeled logic gate is a resistive network, shown in Fig.\ref{fig:cram_basics}(b). 
%$R_1$, $R_2$ and $R_{Out}$ capture the resistances of input and output cells, $In_1$, $In_2$ and $Out$. 
%After this resistive network is established, the output cell is written with -- {\em preset to} -- a known value. 
%Then, a fixed gate-specific voltage $V_{gate}$ is applied at the terminals to guarantee that $Out$ changes state as a function of $In_1$ and $In_2$, according to the truth table of the respective $gate$. 
%This gives rise to electrical currents $I_1$, $I_2$ and $I_{Out}$, respectively, through cells $In_1$, $In_2$ and $Out$.
%The resistances (i.e., logic states) of the input cells modulate $I_{Out}$. As a result, $I_{Out}$ may or may not exceed $I_{crit}$, allowing for the conditional resistance change (hence logic state) of $Out$.
%Boolean gates are performed by presetting or initializing the output cell to a known value and by linking that cell to one or more input cells whose states will modulate the amount of current that passes through the output, where a sufficiently high current can change the cell state. 
%Table~\ref{tbl:and} provides the truth table of a 2-input NOR as a representative example. 

As shown in Fig.1(b), the logic gate operates as a resistive network with cell resistances $R_1$, $R_2$ and $R_{Out}$
corresponding to $In_1$, $In_2$ and $Out$. 
To perform a gate operation, $Out$ is first written with---{\em preset to}---a known value. Applying a gate-specific voltage $V_{gate}$
then drives currents ($I_1$, $I_2$ and $I_{Out}$) through the cells. The resistances (i.e., logic states) of the input cells modulate $I_{Out}$, causing it to conditionally exceed $I_{crit}$ and update the state of $Out$ according to the gate's truth table.
Table~\ref{tbl:and} provides the truth table of a 2-input NOR as a representative example. 

%NVCiM architectures support standard memory read/writes as well as logic operations in memory. Reads involve applying a predetermined read voltage across cell terminals that induces a current significantly lower than $I_{crit}$. The current level is sensed by peripheral sense amplifiers to identify the logic state. Writes operate similarly, but use higher voltages that produce currents above $I_{crit}$ to induce logic state changes. 

%Digital NVCiM designs differ from each other not only by cell technology, but also by the type of supported gate operations and the orientation of computations.
%The range of Boolean operations supported by an NVCiM is a design parameter depending on the optimization targets of the device. 
%Gate operations with varying numbers of inputs are possible, although fan-in is limited by noise margins \cite{zabihi2019spinhall}. The bottom line is that digital NVCiM architectures can support various universal Boolean gates and are functionally complete. %Another limitation 
%Gate operations with different numbers of inputs give rise to {\em instructions}, similar to classic architectures~\cite{leitersdorf_aritpim_2023,resch2022ex}.% \red{please also add a citation to arithPIM here}.  

Digital NVCiM designs vary not only in cell technology, but also in data layout and supported gate operations. While gates with variable number of inputs (fan-in) are realizable, maximum fan-in remains constrained by noise margins \cite{zabihi2019spinhall}. Nevertheless, digital NVCiM architectures support a range of universal Boolean gates, rendering them functionally complete. 
Multi-bit operations spanning standard operand widths and different operand counts form instructions, analogous to conventional architectures ~\cite{leitersdorf_aritpim_2023,resch2022ex}.

\begin{table}[t]
\caption{
Truth table of a universal gate (NOR) with logic 0 preset.
%Truth table of a universal (2-input NOR) gate.
%implemented in MRAM. 
%The output is preset to logic 0. 
%$R_P$ ($R_{AP}$) captures the low (high) resistance state which encodes logic 0 (1).
} 
\vspace{-.2cm}
\label{tbl:and}
% \vshrink{0.2}
\begin{center}
\resizebox{.9\linewidth}{!}{
\begin{tabular}{|ll|c|ll|}
\hline
  In$_1$ ($R_{1}$)  & In$_2$ ($R_{2}$) & Out ($R_{Out}$) & $I_{Out} = I_{1} + I_{2}$ &\\
  \hline 
  0 ($R_{P}$) & 0 ($R_{P}$) & 0 $\rightarrow$ 1 & $I_{00}$ & $> I_{crit}$ \\
 \hline 
 0 ($R_{P}$) & 1 ($R_{AP}$) & 0  & $I_{01}$ & $< I_{crit}$ \\
  \hline
  1 ($R_{AP}$) & 0 ($R_{P}$) & 0 & $I_{10}= I_{01}$ & $ < I_{crit}$ \\
  \hline
  1 ($R_{AP}$) & 1 ($R_{AP}$) & 0 & $I_{11}$ & $< I_{crit}$ \\
  \hline
\end{tabular}
}
%\vspace{.1cm}
%\caption{
%Truth table of a universal, 2-input nand gate implemented in mram ($Output$ preset = 1). 
\vspace{-.85cm}
\end{center}
\end{table}

\noindent{\bf Fundamental Limitations:}
Digital NVCiM designs can orient computation along rows, columns, or bidirectionally~\cite{resch2023pimcity}. Regardless of orientation, however, only a single gate operation can execute at a time along a given row or column. In the column-oriented setup in Fig.\ref{fig:cram_basics}, for example, input and output operands occupy distinct---though not necessarily adjacent---columns within the same row. Because a row can process only one gate operation at a time, all gate evaluations along that row must occur sequentially, creating a NVCiM-specific {\bf serial bottleneck}~\cite{simplermagic,zabihi2018memory,resch2020mouse,resch2023pimcity}.
Conversely, digital NVCiM architectures inherently support Single Instruction Multiple Data (SIMD) execution, as control lines (used to configure the type of computation) span the array orthogonally to the direction of computation. In Fig.\ref{fig:cram_basics}, this enables the identical logic operation to execute simultaneously across matching columns in all rows. While this intrinsic SIMD parallelism helps mitigate the performance impact of the serial bottleneck, it cannot eliminate it. 
In conventional NVCiM, every logic operation requires a preceding preset, consuming a substantial fraction of the total execution time and energy. Consequently, high preset overhead diminishes the effectiveness of SIMD parallelism in compensating for the serial bottleneck.

Furthermore, non‑volatile memory technologies exhibit {\bf limited write endurance}---defined as the maximum number of writes a cell can sustain before permanent error onset---restricting the total permissible number of writes over the array’s lifespan. Output cells, in particular, undergo accelerated degradation due to frequent state updates~\cite{resch2023endure}. Additional writes introduced by preset operations further exacerbate the wear.
Strategically eliminating presets without compromising functional correctness can significantly reduce the effective latency, energy consumption, and write count per logic operation. Specifically, lowering effective gate latency directly alleviates the serial bottleneck; reducing per-gate energy enhances overall energy efficiency, and minimizing writes per cell prolongs hardware endurance.

\section{{\arch: Spatio-Temporal Compiler Optimization}}
\label{sec:design}
%%%%%%%%%%%%%%%%%%%%%%%%%%%%%%%%%%%%
In state-of-the-art digital NVCiM architectures, each logic gate evaluation comprises two distinct steps:
\begin{itemize}
\item[(i)] \textbf{Preset (Pr):} Initializing the target output cell to an operation-specific logic state.
\item[(ii)] \textbf{Thresholding (Th):} Conditionally writing to the output cell to update its state as a function of the input states (Section~II).
\end{itemize}
Although Boolean logic is inherently combinational, this two-step evaluation mechanism imparts sequential semantics to NVCiM, rendering the underlying architecture stateful.

{\em \arch{} is grounded in the observation that the Preset step is not strictly required for functional correctness.} When mapped onto an NVCiM substrate, an application translates into a sequence of logic gates. Mapping an application to NVCiM—referred to as ``compilation'' throughout this paper by analogy to conventional architectures—encompasses both a temporal and a spatial dimension. The temporal dimension dictates the execution order of gate operations required for the computational task. The spatial dimension, conversely, governs data layout and physical gate locations within the memory array.
Consider a sequence of gate operations $G_i$ and $G_{i+1}$ in program order, with respective {\em Preset} and {\em Thresholding} steps 
$Pr_i$, $Pr_{i+1}$, $Th_i$ and $Th_{i+1}$. Intuitively, $Pr_{i+1}$ can be safely eliminated if $G_{i}$ leaves the output cell in the preset state required by $G_{i+1}$. However, this condition holds only if the outputs of both $G_i$ and $G_{i+1}$	
  map to the exact same physical memory cell. Consequently, tracking output values and preset requirements along execution order alone (the temporal dimension) is insufficient to eliminate redundant presets; spatial cell allocation within the array is equally critical. This spatio-temporal coupling distinguishes preset optimization in NVCiM from conventional compilation and logic synthesis flows.

{\em Furthermore, \arch{} extends beyond simple redundant preset elimination.} By leveraging formal spatio-temporal analysis across gate sequences, \arch{} eliminates presets while strictly {\em preserving end-to-end computational correctness} in the application context. Beyond per-gate preset removal, \arch{} identifies opportunities to strategically coalesce or eliminate subsequent {\em Thresholding} steps once presets are removed. In contrast, state-of-the-art NVCiM compilers overlook these opportunities by conservatively treating every single preset as indispensable.

\subsection{Formal Optimization Rules}
\label{sec:rules}

\arch{} categorizes gate operations into preset-0 and preset-1 types based on their required initial state. Depending on the specific NVCiM architecture, the required preset value for a given logic function may vary, even though the underlying operating principle of gate realization remains unchanged. For instance, a NOR operation functions as a preset-1 gate in the MAGIC architecture~\cite{magic}, whereas it serves as a preset-0 gate in MOUSE~\cite{resch2020mouse}—similar to the example in Table~\ref{tbl:and}. Table~\ref{tab:presetcategory} summarizes this categorization for basic Boolean logic operations (and their corresponding instructions) considering the MRAM-based NVCiM baseline detailed in Section~\ref{sec:bkg}.

\begin{table}[t]
    \caption{Categorization of gate operations by preset value. $MAJ$ ($\overline{MAJ}$) is the {\em majority} gate that outputs true (false) if more than half of its inputs are true (false).}
    \vspace{-.1cm}
    \centering
    \footnotesize
    \begin{tabular}{|c|c|}
    \hline
         Preset-0&Preset-1 \\
         \hline
         NOT, NAND, NOR, $\overline{\mathrm{MAJ}}$ & COPY, AND, OR, MAJ \\
         \hline
    \end{tabular}
    \label{tab:presetcategory}
    \vspace{-0.5cm}
\end{table}

Due to MTJ switching physics, the following invariant holds for all preset-0 logic gates~\cite{zabihi2019spinhall} %\red{[can we add one or a few device physics references here, could be from JP?]}
: the \textit{Thresholding} step can only conditionally induce a $0 \rightarrow 1$  state transition. In other words, if the designated output cell holds logic 1 prior to \textit{Thresholding}, it inherently preserves its value; non-volatile device physics prevents a $1 \rightarrow 0$  transition under a preset-0 configuration. Consequently, if an output cell is evaluated to $Out_i=1$ by a gate $G_{i}$, the post-thresholding state of a subsequent gate $G_{i+1}$ in program order sharing the same physical output cell ($Out$) will remain $Out_{i+1} = 1$. This behavior aligns directly with NVCiM execution semantics: conditional computation requires a specific initial preset state, without which the target cell preserves its existing state.

In summary, the only valid state transition for a preset-0 gate is $0 \rightarrow 1$. Symmetrically, for a preset-1 gate, the only valid state transition is $1 \rightarrow 0$. We formalize this property as follows:

{
\vspace{.2cm}
\noindent\fbox{
%\vspace{.1cm}
\begin{minipage}[c]{.97\linewidth}
%\begin{itemize}
%\item 
\bf{Preset-0 Rule} applies if $\mathbf{G_{i+1}}$ is a preset-0 gate:\\
%\hfill $\mathbf{G_{i+1}: Out_{i+1} = Out_{i}~~|~~Th_{i+1}(In_1, In_2,...)}$\\
\indent $\mathbf{Out_{i+1} = Out_{i}~~|~~Th_{i+1}(In_1, In_2,...)}$\\
%\item 
\bf{Preset-1 Rule} applies if $\mathbf{G_{i+1}}$ is a preset-1 gate:\\
\hfill $\mathbf{Out_{i+1} = Out_{i}~~\&~~Th_{i+1}(In_1, In_2,...)}$
%\end{itemize}
\end{minipage}\hfill
%\vspace{.1cm}
}
\vspace{.1cm}
}

%$G_i$ here corresponds to a gate operation (preceding $G_{i+1}$ in program order), which shares the same output cell as $G_{i+1}$. We can think of $G_i$ as the last writer to the respective output cell $Out$ before $G_{i+1}$. The very same observations, and hence the rules, apply beyond the gate context. Therefore, we can think of $G_i$ as representing any sequence of gate operations or computations that share the end output cell with $G_{i++}$ and that represents the last writer to the target output cell. Accordingly, \arch\ transitively applies these two rules in eliminating presets. In the end, these two rules describe the interaction between the end result of a prior operation or operations, and a subsequent gate without a preset through the shared output cell. 

Here, $G_i$ denotes a gate operation preceding $G_{i+1}$ in program order that shares the same target output cell $Out$, effectively serving as the most recent writer (``last writer'') to $Out$ before $G_{i+1}$. These observations---and the resulting rules---extend beyond individual gate pairs. Specifically, $G_{i}$ can represent any preceding sequence of computations whose final output is written to $Out$ prior to $G_{i+1}$. Accordingly, \arch{} transitively applies these rules across multi-gate sequences during preset elimination. Fundamentally, these rules govern the spatio-temporal interaction between the final result of prior operations (as captured by $Out_{i}$) and a subsequent gate without preset (as captured by $Th_{i+1}$) through their shared output cell. 

\subsection{\arch-Enabled Code Compaction: Microscopic View}
%Impact on Instruction Sequence Length}
%Instruction Count
%Minimizing Preset-induced Overheads at the Gate-level}

\begin{figure*}[thp]
\centering
\includegraphics[scale=1,height=5.3cm]{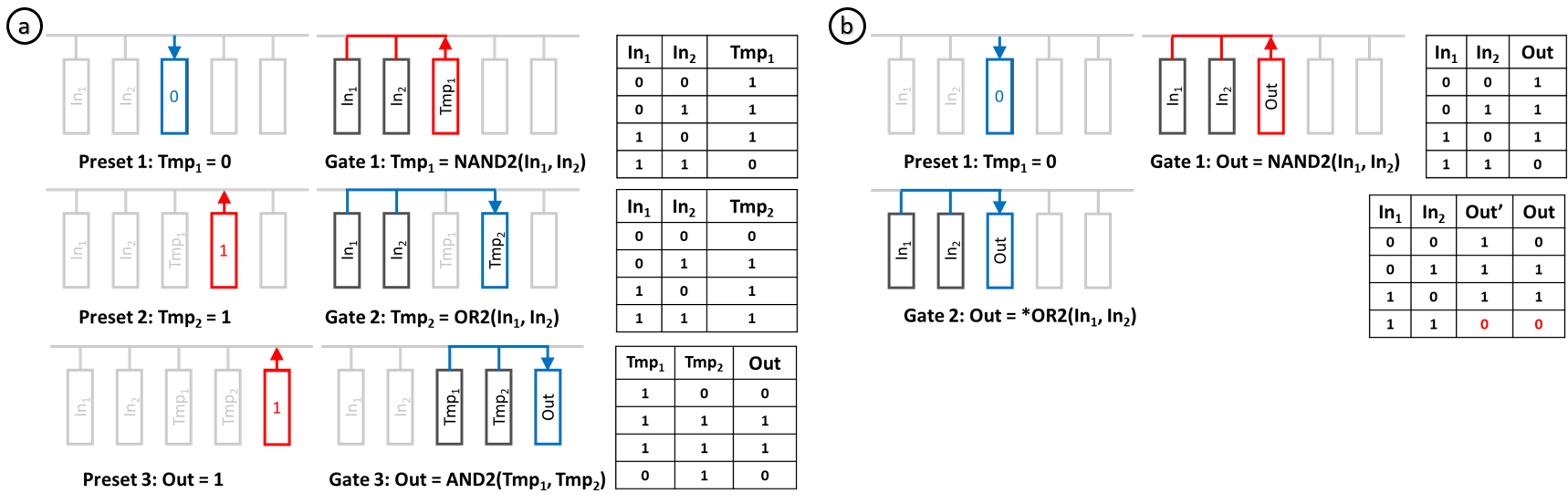}    
\caption
%{XOR with truth table: a) Standard XOR diagram with 3 gates, paired truth table changes across 3 cells. b) Optimized XOR with 2 gates with %evolution across one cell. Green squares indicate input cells, yellow are scratch or output cells. Circled numbers indicate gate order. Dotted %gate is an implied operation. Solid gate with dashed left side has no preset.
%\red{can we sub A,B with $In_1$ and $In_2$?}
%}
{XOR mechanics considering the implementation from Table~\ref{tbl:xor}: a) State-of-the-art NVCiM implementation featuring three two-step (preset + thresholding) gates. For each gate, the Preset step sets the output cell highlighted in blue/red to 0/1. An arrow indicates if the operation pulls the cell down/up. The gate's threshold step connects the inputs and outputs, with the current path highlighted in blue/red and a down/up arrow indicating when the gate pulls the output to 0/1. b) \arch\ XOR featuring two gates, one of which has its preset eliminated and labeled with an asterisk. Truth tables indicate the Boolean operation completed by each step, with the optimized gate's table showing the modified behavior of the Out cell.
%Standard XOR diagram with 3 gates, paired truth table changes across 3 cells. b) Optimized XOR with 2 gates with evolution across one cell. 
%Green squares indicate input cells; yellow, scratch (to hold temporary values) or output cells. Circled numbers give the order of gate evaluations. Each gate is labeled with where its output goes. Dotted gates are (implied, hence) eliminated operations. Solid gates with dashed left sides have no preset. Gate names with an asterisk (*) indicate \arch-optimized gates.
%\red{[Out color in b?]}
%\red{can we sub A,B with $In_1$ and $In_2$?}
}
\label{fig:combxortable}
\end{figure*}

{\em When mapped onto a digital NVCiM architecture, an application reduces to a sequence of primitive logic gates, where multi-bit, multi-operand variants of these gates correspond to instructions. Consequently, any reduction in total gate count directly yields a more compact instruction stream.}
Digital NVCiM architectures implement Boolean functions via a two-step evaluation process comprising presetting and thresholding, and a broad spectrum of logic gates remains realizable.
A simple inspection of a function's truth table reveals whether a direct NVCiM implementation is feasible. By construction, ordered traversal of truth-table rows corresponds to a monotonic change in equivalent input resistance, as input logic states are physically encoded as resistance values. This causes the resulting output current ($I_
{Out}$) to decrease monotonically. Consequently, NVCiM can directly realize any Boolean function whose output column exhibits contiguous blocks of 0s and 1s (e.g., the NOR operation in Table~\ref{tbl:and}). This monotonic response partitions $I_{Out}$ into two distinct regimes relative to a critical switching current ($I_{crit}$): an $I_{Out} > I_{crit}$ regime that enforces an output state transition during thresholding, and an $I_{Out} < I_{crit}$ regime that retains the initial preset value.

This monotonicity condition does not hold for non-linearly separable functions such as XOR or XNOR. Consequently, state-of-the-art NVCiM architectures cannot execute XOR as a single-gate operation~\cite{felix2018, chowdhury_spintronic_2019}. Nevertheless, XOR and XNOR represent vital computational primitives---applications ranging from pattern matching in bioinformatics to fundamental arithmetic units for multiply-accumulate (MAC) operations in machine learning workloads.
Consider a 2-input XOR operation as an illustrative example. Underlying device physics dictate that $I_{00} > I_{01} = I_{10} > I_{11}$. Correct XOR evaluation requires the output cell to maintain the same state for the 00 and 11 input patterns, while undergoing a state transition for 01 and 10 (or vice versa). However, because 
$I_{01}$ and $I_{10}$ lie strictly between $I_{00}$ and $I_{11}$, no choice of preset state or critical switching current ($I_{crit}$) can isolate {$I_{00}$ and $I_{11}$} from $I_{01}$ and $I_{10}$. Although single-gate realization is impossible, digital NVCiM architectures can synthesize the XOR function using a multi-gate sequence of universal primitives, as detailed in Table~\ref{tbl:xor}. \arch-enabled code compaction can be particularly advantageous in this context.

\begin{table}[t]
\caption{NVCiM XOR implementation combining three gates. 
%NAND is a preset-0 gate; OR and AND, preset-1 gates.
} 
% \vshrink{0.2}
\begin{center}
\resizebox{.85\linewidth}{!}{
\begin{tabular}{|ll|c|c|c|}
\hline
& & S$_1$= & S$_2$= & $Out$=\\
  In$_1$   & In$_2$  & NAND(In$_1$,In$_2$)  & OR(In$_1$,In$_2$) & AND(S$_1$,S$_2$)\\
  \hline 
  0  & 0  & 1  & 0 & 0 \\
 \hline 
 0  & 1  & 1  & 1 & 1\\
  \hline
  1  & 0  & 1 & 1 & 1\\
  \hline
  1  & 1  & 0 & 1 & 0\\
  \hline
\end{tabular}
}
%\vspace{.1cm}
%\caption{
%Truth table of a universal, 2-input nand gate implemented in mram ($Output$ preset = 1). 
\label{tbl:xor}
\vspace{-.9cm}
\end{center}
\end{table}

Fig.\ref{fig:combxortable} details the execution mechanics of the XOR implementation from Table~\ref{tbl:xor}, contrasting conventional state-of-the-art NVCiM (Fig.\ref{fig:combxortable}(a)) with \arch{} (Fig.\ref{fig:combxortable}(b)).
In the state-of-the-art baseline (Fig.\ref{fig:combxortable}(a)), the NAND and OR gates require distinct output cells, which subsequently serve as inputs to a third AND operation (Table~\ref{tbl:xor}). In contrast, \arch{} applies the {\bf Preset-1 Rule} (Section~\ref{sec:rules}) to the AND step. As a result, executing an ``un-preset'' OR operation (*OR) in the cell shared with NAND eliminates the need for the third gate evaluation entirely, as shown in Fig.\ref{fig:combxortable}(b). 
Although the un-preset *OR operation does not strictly match the standard OR truth table (due to the mismatch for $In_1In_2=11$), it replicates the combined functionality of an OR operation followed by an AND operation in conventional, preset-based NVCiM.
Following this same principle, \arch{} compresses XNOR into an AND operation followed by a *NOR operation, which generates the output sequence $\{1,0,0,1\}$ matching the standard XNOR truth table.
This representative example demonstrates that \arch{} optimizations---by propagating {\bf Preset-0 Rule} and {\bf Preset-1 Rule} (Section~\ref{sec:rules}) along dependent operation chains---can eliminate not only individual presets, but entire gate evaluations. This directly yields more compact instruction sequences, driving higher performance, improved energy efficiency, and reduced wear.

\begin{figure}[tbp]
\centerline{\includegraphics[scale=0.4]{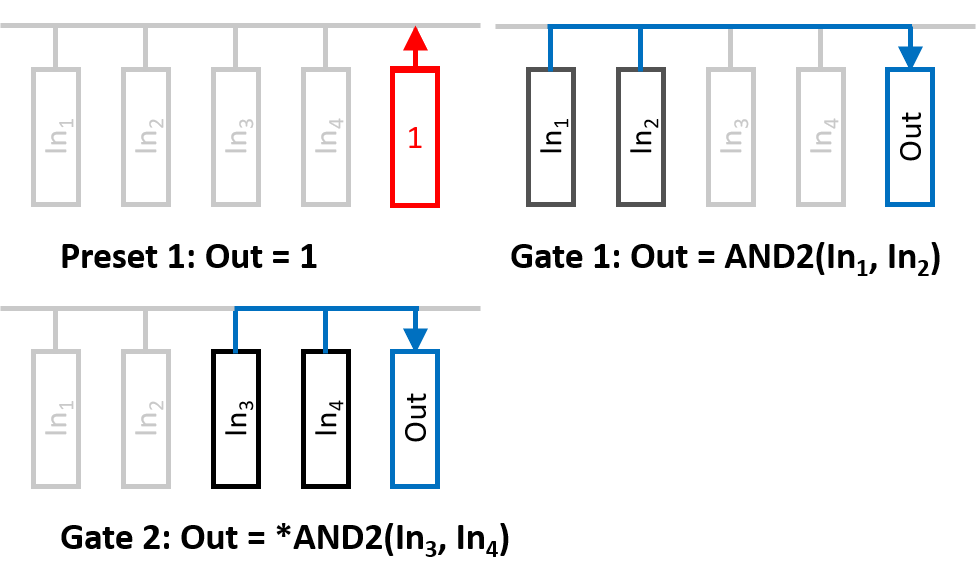}}
\vspace{-.1cm}
\caption{Optimized synthesis of AND4 from AND2 gates. Preset step sets the output cell highlighted in blue/red to 0/1. The current path is highlighted in blue/red with a down/up arrow indicating if the operation pulls the cell down/up.
%\red{[once finalized, let us make sure that the fig. sizes (of similar format as in here) are balanced]}
}
\vspace{-0.1cm}
\label{fig:and4}
\end{figure}

\begin{figure}[tbp]
\centerline{\includegraphics[scale=0.4]{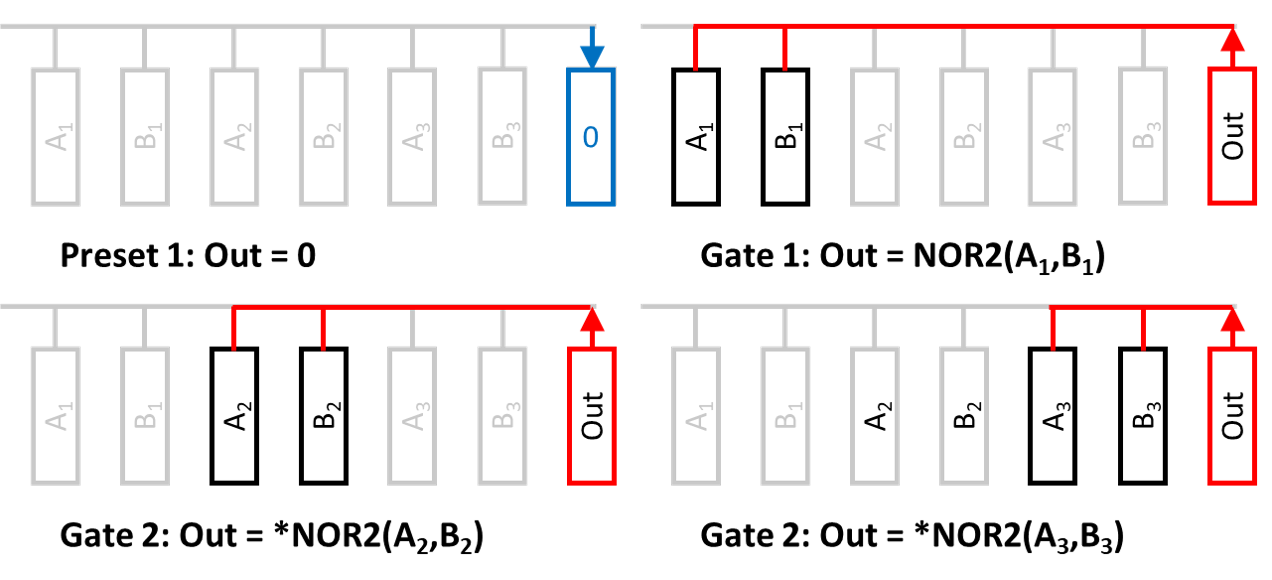}}
\vspace{-.1cm}
\caption{Optimized synthesis of C $\leftarrow C \lor (A_i \land B_i)$, using inverted versions of vectors A, B. Preset step sets the output cell highlighted in blue/red to 0/1. The current path is highlighted in blue/red with a down/up arrow indicating if the operation pulls the cell down/up.
}
\vspace{-0.1cm}
\label{fig:norchain}
\end{figure}

Although physical and electrical constraints bound the maximum fan-in, a broad spectrum of logic functions remains realizable by decomposing high-fan-in gates into sequences of natively supported lower-fan-in NVCiM primitives. \arch{}-driven code compaction significantly optimizes this decomposition.
For example, synthesizing a 4-input AND operation (AND4) using conventional 2-input gates (AND2) typically requires three separate preset-based gate evaluations arranged in a reduction tree. In contrast, as depicted in Fig.\ref{fig:and4} %\red{  [can we add a small example figure, using a similar format to the XOR fig.?]}, 
%\blue{AND4 optimized variant only, or also the original? Include truth tables?}
\arch{} evaluates an AND4 by cascading an initial AND2 with an un-preset *AND2 operation in the same shared output cell.
This principle generalizes to arbitrarily wide AND gate operations beyond AND4. Under \arch{}, the total gate count scales linearly with the number of inputs without requiring extra combining gates for intermediate results—unlike conventional NVCiM architectures. Consequently, the code compaction ratio increases with the emulated gate fan-in.

Section~\ref{sec:app} presents additional examples of \arch{}-optimized computational primitives. Supporting a wider variety of gate types yields substantial savings and enables compact implementations of complex blocks, such as the full adder in Fig.~\ref{fig:adder_comp}. For instance, whereas restricting implementation to two-input NOR gates requires 9 gates for a full adder, \arch\ produces a mixed-gate alternative that achieves a $3\times$ speedup. Furthermore, diverse gate mixes allow for direct, highly efficient realizations of complex circuits, such as the comparator in Fig.~\ref{fig:compar_circ}.

\subsection{\arch-Enabled Code Compaction: Macroscopic View}
%Impact on Instruction Sequence Length}
%Instruction Count
%Minimizing Preset-induced Overheads at the Gate-level}

In state-of-the-art NVCiM architectures, a memory cell cannot simultaneously serve as both an input and an output within a single computational step due to conflicting terminal voltages. This constraint, as well, introduces a serialization bottleneck and complicates the NVCiM implementation of essential computational primitives such as in-place accumulation. 

Consider the binary vector conflict-check operation 
$$C \leftarrow C \lor (A_i \land B_i)~~~~~~~i \in \{1,2,…,n\}$$ 
as a representative example, where $A_i$ and $B_i$ denote elements of binary vectors of length $n$, 
%with $i \in \{1,2,…,n\}$, 
and $C$ is an accumulator flag indicating whether any corresponding pair satisfies $A_i = B_i =1$. 
In state-of-the-art NVCiM architectures, the OR operation evaluating $A_iB_i \lor C$ 
%$A_iB_i = A_i \land B_i$ 
must output to a temporary intermediate cell because $C$ cannot supply its own input current while simultaneously acting as an output cell. A dedicated COPY operation is then required to update $C$. Consequently, processing each vector element requires three full two-step gate evaluations.
\arch{} can eliminate this overhead through two distinct optimization strategies:
\begin{itemize}
\item Inverted Input Domain: If vectors A and B are stored in inverted format ($\overline A_i$ and $\overline B_i$), a valid \arch{} sequence consists of an initial NOR operation followed by iterative un-preset *NOR operations, as depicted in Fig.\ref{fig:norchain}. This approach reduces execution toward a single thresholding step per index. In NVCiM architectures supporting dual-axis row/column computation, vectors can be inverted rapidly along one axis prior to iteration along the other, though the energy cost of inversion still applies.
\item Default Input Domain: When inputs are in true-form, $C \leftarrow C \lor (A_i \land B_i)$ can be transformed, using De Morgan's laws, into $\overline C \leftarrow \overline C \land \overline {A_i \land B_i}$. Each iteration then executes a NAND operation followed by an un-preset *COPY step (which acts as an implicit AND on $\overline C$). This reduces per-index execution to two gate evaluations---one of which is preset-free---culminating in a single final inversion (INV or NOT) at the end of the vector.
\end{itemize}

Ultimately, \arch{} enables persistent state accumulation across iterations by structuring logic mappings such that implicit gate operations preserve accumulator state while bypassing the electrical constraints of thresholding.

\subsection{Putting It All Together}

The NVCiM programming model is analogous to GPU programming with SIMD-like thread blocks, where the primary optimization objective is instruction (and consequently, gate) scheduling across both time and space. In this context, instructions represent multi-operand, multi-bit extensions of basic gate operations. Existing NVCiM tools such as PyPiM~\cite{pypim24} and CiNM~\cite{CINM25}, support application mapping by offering specialized libraries of optimized computational primitives. However, unlike \arch, neither toolchain is preset-aware.

\arch\ expands the design space of NVCiM compilers by incorporating preset awareness, unlocking optimization opportunities across performance, energy efficiency, and endurance. 
Specifically, \arch\ serves two primary roles within the NVCiM optimization flow:

\begin{itemize}
\item {\em Pre-Scheduling Logic Discovery:} Identifies a broad set of logic operations (and combinations thereof) prior to final spatio-temporal code scheduling on the target NVCiM substrate. This offers early guidance to compilation tools, avoiding suboptimal results caused by constrained initial search spaces.  
%or of post-synthesis logic restructuring.
%This guides downstream tools and avoids missed optimizations stemming from overly restricted options or 
%complex gate-level re-evaluations.
\item {\em Architectural Co-Design Guidance:} Informs key design choices during NVCiM development. \arch\ evaluates high-impact computational primitives and preset assignments for target workloads, preventing performance bottlenecks caused by mismatches between hardware primitives and workload demands.
\end{itemize}

\begin{figure}[t]
    \centering
    \includegraphics[width=1.0\linewidth]{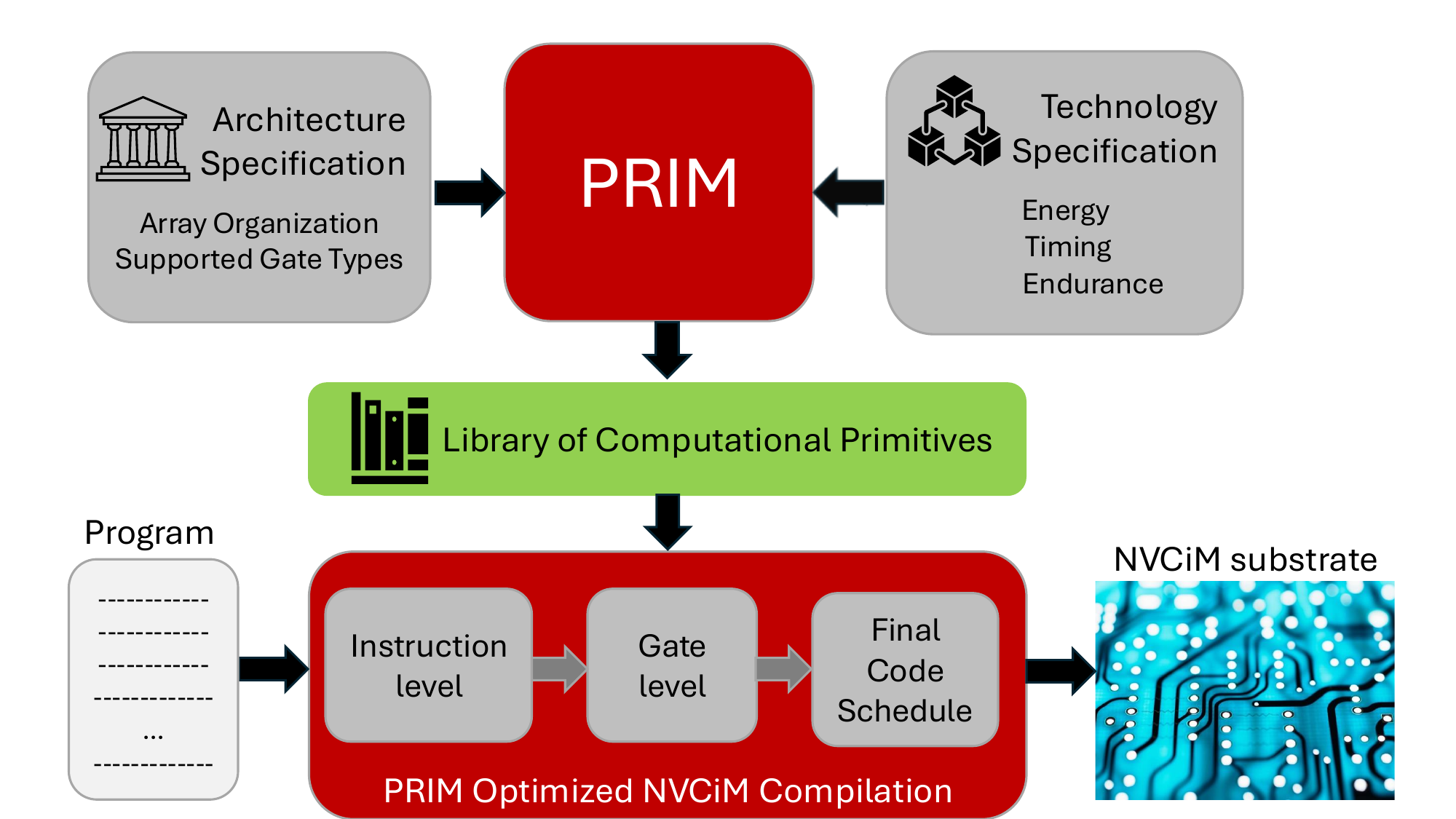}
    \vspace{-.3cm}
    \caption{\arch{}-optimized compilation flow for digital NVCiM.}
    \label{fig:piat}
    \vspace{-.3cm}
\end{figure}

%%% Flow

At the core of \arch\ optimizations is the transitive application of the Preset-0 and Preset-1 rules introduced in Section~\ref{sec:design}. \arch\ applies these rules across multiple scales. It targets individual computational primitives, and can chain blocks together to form complex structures such as application-specific kernels—up and applications. As outlined in Fig.~\ref{fig:piat}, a key application of these preset-aware optimizations is the synthesis of optimized computational primitive libraries~\cite{pypim24}. \arch{}-enhanced NVCiM compilers can parse high-level programs and directly map operations (e.g., conditionals, arithmetic, and selection) to optimized primitives tailored to specific performance, energy, or endurance targets. 
%Once these custom libraries are constructed, general compilation and synthesis pipelines can directly leverage \arch{}-optimized primitives. 
Backend frameworks such as SIMPLE~\cite{simplermagic} can then generate the final NVCiM code stream.

%The analysis of optimized sequences or libraries for a  device feed into a generation stage, stage (3) in Fig.\ref{fig:flowchart}. 
%shows where this process can be inserted in the toolchain for application mapping. 
%Preset optimization 
%\arch\ can use architecture-specific lists of supported NVCiM gates to build a library of blocks that maps to computational primitives (+,-,$\times$,etc.) used by higher-level tools like CiNM without modifying them to be preset-aware. \arch\ can also  
%can be used by higher-level interfaces such as CiNM, or 
%generate the equivalent of standard cells or lookup tables (LUT) for input/output combinations, along with delay and energy costs.
%that can be synthesized more optimally by preset optimization, which can be used 
%Logical synthesis tools such as ABC \cite{mishchenko2007abc} can directly use these as part of existing non-preset aware methods. 

%%% Algorithm

To construct an optimized library of computational primitives, \arch\ decomposes functional blocks into truth tables. Algorithm~\ref{alg:blockopt} (Section~\ref{sec:app}) details a keyhole optimization-based approach that enforces the Preset-0 and Preset-1 rules from Section~\ref{sec:rules}: By working backward from the target truth table and reversing gate operations, \arch\ determines whether a valid path to a preset state exists, deriving an optimized sequence in reverse order. If a such a chain of preset-free operations is not found, \arch\ identifies valid partial chains and stitches them together. While lightweight heuristics—such as greedy algorithms that match gates to the largest portion of a truth table—can provide initial candidate sequences, they frequently miss globally optimal solutions.

Although the unconstrained combination of operations and inputs could grow excessively large, \arch\ prunes search paths that perform worse than baseline preset-aware sequences. % (as illustrated in Fig.~\ref{fig:tree}, Section~\ref{sec:app}). 
Because NVCiM architectures are inherently restricted to a compact set of operations and bounded fan-in counts, the search space for \arch-based optimization remains strictly bounded. Furthermore, since physical hardware implementation parameters are fixed, many of these evaluations can be pre-computed during library generation.

\section{{Evaluation Setup}}
\label{sec:eval_setup}
%NVCiM relies on its integration of memory and processing to circumvent bandwidth limitations, and it relies on internal parallelism to compensate for its poor sequential performance with massive throughput. The inverse relationship between performance and gate count strongly encourages circuits that have the fewest gates, as these are the base units of execution.
\begin{table}[t]
    \centering
    \begin{tabular}{|c|c|c|}
    \hline
    Parameter & STT\cite{cilasun2024ecc} & SOT/SHE\cite{zabihi2019spinhall} \\ [0.5ex]
 \hline\hline
 $R_P$ $(k\Omega)$ & 3.15 & 253.97\\ 
 %%%Where did error come from?
 %$R_{AP}$ $(k\Omega)$ &  76.39 & 507.94\\
 $R_{AP}$ $(k\Omega)$ &  7.34 & 507.94\\
 $R_{SHE}$ $(k\Omega)$ & - &64\\
 $I_{crit}$ $(\mu A)$ & 50& 3\\
 $t_{sw}$ $(ns)$  & 1 &  1 \\
 \hline 
 $t_{Clk}$ $(ns)$ & 3 &3\\[1ex]
\hline
    \end{tabular}
    \vspace{.1cm}
    \caption{MRAM technology parameters.}
    \label{tab:cellparams}
    \vspace{-.8cm}
\end{table}
%\subsection{Instruction Improvements}
%The base design of the NVCiM maps a preset and gate operation to one instruction. Per Table \ref{tab:cellparams}, the model design has a 3ns period, which is composed of two 1ns periods for output cell switching, and 1ns for the required propagation of signals and settling periods. The most straightforward implementation of preset removal would suppress the preset pulse without changing anything else in the system. 

%In many cases, preset optimizations will still improve performance due to a decrease in raw gate count. This can leave many instruction cycles 2/3 idle. A more aggressive architecture could coalesce neighboring gates into the same instruction period, or instruction encodings can be changed to create paired instructions. Ideally, an NVCiM would be able to support a wide range of gate operations and also be able to arbitrarily combine them, to find the most ideal combinations. However, implementations can be limited in the number of gate voltages they support\cite{cilasun2024ecc}, and complete freedom in encoding combinations of gates with and without presets can increase complexity of the instructions or the sequencing hardware for an array.

%\subsection
\noindent{\bf Simulation Framework and Benchmarks:}
We use a cycle-accurate instruction-level CiM simulator 
%in the C programming language 
to evaluate the impact of preset optimizations at the functional block level (considering \texttt{XOR}, Full-Adder, multiplexer \texttt{MUX}, and magnitude comparator \texttt{COMP} from Sections~\ref{sec:design} and~\ref{sec:app}) as well as
%on 
more complex computational kernels (a 4-bit Matrix-Vector Multiplication, \texttt{VMUL}, in a standard 1024$\times$1024 array using adders based on Fig.\ref{fig:adder_comp}b or \ref{fig:adder_comp}d). 
For the full-adder we consider Fig.\ref{fig:adder_comp} (a) as a baseline single gate-type adder, (b) as a multiple-gate adder--\texttt{ADD\_MIX}, {and analyze their \arch{}-optimized versions: \texttt{ADD\_NOR} and \texttt{ADD\_OP}.} 
The simulator is augmented by a performance model to estimate time, energy and endurance. 
%
%We developed a Matrix-Vector Multiplication kernel to provide a simple kernel that was still composed of multiple functional blocks. This performs 
%MVM kernel performs binary multiplication of a matrix of randomized 4-bit values with a vector of 4-bit elements
%. In order to fit 
%in a standard 1024$\times$1024 array. %, this set of data and the reserved regions for the arithmetic blocks leads to a length of 59 elements. 
%The kernel uses repeated ripple-carry adders using an adder based on Fig.\ref{fig:adder_comp}c or \ref{fig:adder_comp}d. 
%A single multiplication is simulated and the time, energy, and write counts are recorded.

For larger-scale end-to-end evaluation, we model a classical Hopfield
Network~\cite{hopfield1982neural}, \texttt{NN},  consisting of 128 fully connected neurons with
4-bit weights mapped to a $1024 \times 1024$ NVCiM array. Each iteration
comprises matrix vector multiplication (\texttt{DOT}), neuron update selection (\texttt{SEL}), and state broadcasting. The runtime-dominant  matrix vector multiplication phase performs dot products via two's complement ripple-carry adders with XNOR-based conditional operand negation and 
%The MAC uses 
an 11-bit accumulator.
The update selection phase scans binary vectors for updated neuron states before broadcasting them to prepare the next iteration. We compare a standard-instruction baseline (\texttt{ADD\_MIX}) %Fig.~\ref{fig:adder_comp}c) 
against a preset-optimized variant that replaces XNOR and full-adder blocks with preset-optimized sequences. The update step performs search for non-zero entries in bit-vectors by using long sequences of *NAND gates.

\noindent{\bf Technology Parameters, Timing, Energy, Endurance:} As given in Table~\ref{tab:cellparams}, we consider two representative MTJ technologies: Spin Transfer Torque (STT) and Spin Orbit Torque (SOT)/Spin Hall Effect (SHE).
%an STT similar to modern cells, and projections for an emerging SOT/SHE technology. 
The cell parameters in Table~\ref{tab:cellparams} synthesize empirical MRAM demonstrations and industry technology roadmaps. While STT-MRAM is already commercially deployed, cutting-edge figures of merit—such as sub-nanosecond switching ($250$---$500~\text{ps}$)~\cite{nguyen_recent_2024}, high write endurance ($10^{14}$~\cite{mram_endure} to $>10^{15}$ cycles~\cite{nguyen_recent_2024}), low write energies (tens of $\text{fJ/bit}$)~\cite{nguyen_recent_2024}, and elevated Tunneling Magnetoresistance (TMR) ratios ($\ge 300\%$ at room temperature)~\cite{ICHINOSE2023170750}—have typically been demonstrated in distinct device variants rather than a single monolithic process. Higher TMR directly improves read and logic-gate operating margins, mitigating a key constraint of MRAM relative to other non-volatile memories. Although co-optimizing all metrics within a single commercial process remains an active fabrication goal, continued scaling along industry projections supports the parameter configuration in Table~\ref{tab:cellparams}.
%
%The evaluation performs a sweep across the latency and resistance parameters given for STT and SHE in Table \ref{tab:cellparams}, over the baseline and optimized kernel variations. %Latencies measured in three scenarios: 
%\blue{***Remove preset-only and delay discussion for space?***}

%\noindent{\bf Timing Model:}
The latency of a standard operation encompasses both preset and thresholding
steps. Each step applies a voltage pulse lasting $t_{\text{sw}}$ ($1~\text{ns}$), with an additional $1~\text{ns}$ allocated for ancillary array activity and signal propagation. Output cell initialization scales linearly with the number of cells and is fully accounted for in latency, as repeated invocations of a functional block require re-initialization. Additionally, signal inversions are conservatively counted as full gate operations, assuming inputs originate from prior operations and require explicit negation.
While certain architectures support simultaneous multi-cell writes to apply presets in parallel, physical memory arrays face peak bulk-preset current constraints that limit parallel initialization~\cite{cilasun2021spiking}. Because hiding preset latency via large bulk writes requires substantial current overhead in high-throughput arrays, we conservatively assume a 1:1 ratio between cell initializations and preset latency cycles.

We experiment with three configurations:
{\em Default} features no optimization and each instruction takes 3ns. 
%use a 3ns period, 
{\em Preset/Delay} uses preset-less instructions that take 1.5ns, and {\em Preset Only} {removes the preset from optimized instructions, but conservatively keeps the standard 3ns instruction latency.}
%\red{REMOVED for now; what I meant was semantic explanation of the configurations, not how they'd be implemented.}
%\blue{
%The default scenario of paired preset and thresholds is similar to MOUSE\cite{resch2020mouse}, and the two modified scenarios envision an additional bit flag that can skip the preset step. Other architectures may have the ability to dispatch presets independently\cite{simplermagic}, and would require no change to take advantage of \arch.}
%[please double check and revise for clarity. Not clear what the diff of Default and Preset/Delay is based on this description, as well as what Preset Only serves.]
%\red{keeps latency constant}. 
%The simulated NVCiM array is initialized and then executes the kernel, which iteratively converges towards a stored pattern. The time taken for this process, the energy taken based on average gate consumption, and the write counts per cell were recorded. 

%\subsection{Calculating Gate Energy}
%\noindent{\bf Energy Model:}
The energy cost of an instruction can include many second-order effects. %For the purpose of providing an estimate, 
We calculate the sum of the preset and thresholding steps---which represent the dominant components--for the in-array energy cost of each instruction.  Overheads induced by instruction fetch and sequencing scale directly with instruction count and are inherently reduced by preset optimizations. Specifically, we calculate gate energy using parameters for STT~\cite{cilasun2024ecc} and SOT/SHE~\cite{zabihi2019spinhall} as follows:
\begin{equation}
    E_{op}=\frac{V_{write}^2}{R_{out}}*t_{sw} + \frac{V_{gate}^2}{R_{eq}}*t_{sw}
\label{eq:e_op}
\end{equation}
The first term models preset energy, where $V_{\text{write}}$ represents the write voltage and $R_{\text{out}}$ is the resistance of either the STT cell~\cite{cilasun2024ecc} or the SOT/SHE material~\cite{zabihi2019spinhall}. The second term captures thresholding energy, where $V_{\text{gate}}$ is the gate voltage and $R_{\text{eq}}$ is the equivalent resistance of the parallel input cells in series with the output cell or SHE channel. Both technologies assume a cell switching time ($t_{\text{sw}}$) of $1~\text{ns}$.
STT energy modeling is more complex because output cell resistance varies during preset and thresholding steps, requiring higher voltages for Preset-1 operations than their inverse counterparts. Additionally, preset write voltages differ because the critical switching current ($I_{\text{crit}}$) is reached at a lower voltage when writing `1' to a cell in the parallel state ($R_{\text{P}}$), allowing Preset-1 instructions to use a lower-energy preset. 

%\noindent{\bf Endurance Model:}
%MTJ cell endurance is a significant concern. 
We evaluate the impact of cell endurance for three configurations---{\em Low} $(10^{12})$, {\em Medium} $(10^{14})$, and {\em High} $(10^{16})$ ---considering different MRAM technologies\cite{resch2023endure,cilasun2024ecc,unlimited_kan,wang_scaling}. Each configuration assumes a different value for the maximum number of writes a cell can endure before the onset of permanent errors, which conservatively demarcates the lifetime of an array.
We assume that a gate or kernel is run continuously
%, with the first metric being 
until {\em Time to First Failure (TTFF)}, which is calculated for when the first cell in an array reaches its projected lifetime limit.

\section{{Evaluation}}
\label{sec:eval}
\begin{figure}[t]
    \centering
    \includegraphics[width=1.0\linewidth]{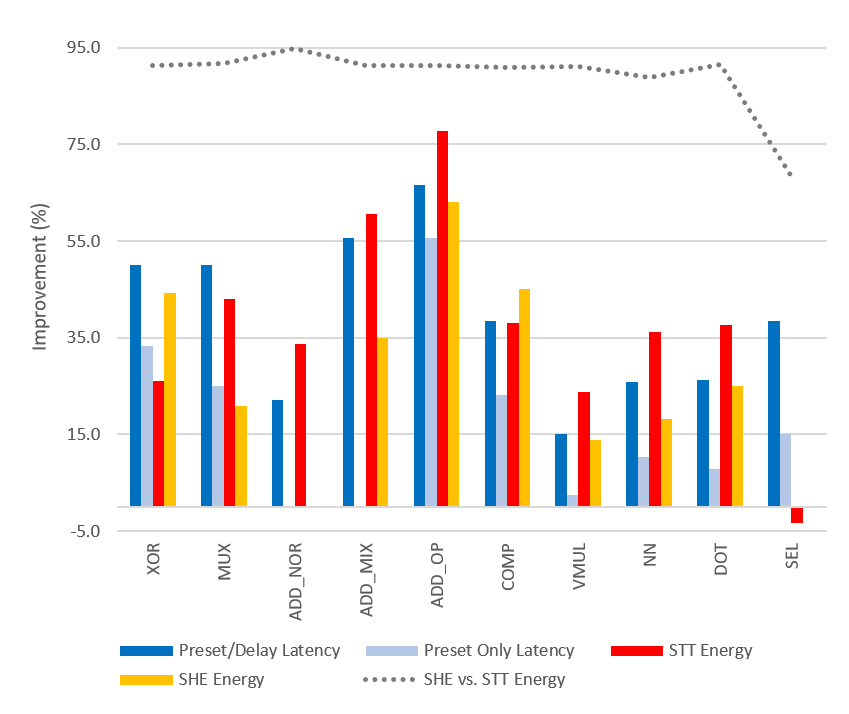}
    \vspace{-.8cm}
    \caption{Percentage improvement using \arch{}. 
    \texttt{ADD\_MIX} has no Preset Only value as it is a secondary adder baseline with no preset optimization. }
    \label{fig:ratio_graph}
    \vspace{-.5cm}
\end{figure}

%We evaluate the proposed preset optimizations at the functional block level, with a matrix-vector kernel, and a 
%and at the level of an application mapping to an NVCIM, with 
%Neural Network kernel composed of many functional blocks with a representative data layout and data flow. 
%The primitives XOR, MUX, and comparator(COMP) are evaluated for latency improvement versus the standard implementation and energy reduction across STT and SHE memory types. The full-adder case is evaluated across variants illustrated in Fig.\ref{fig:adder_comp}. Using (a) as a baseline, the configurations in (b)-(d) are covered by ADD\_NOR, ADD\_MIX, and ADD\_OP, respectively. The two computational kernels are covered by the VMUL and NN cases. Fig.\ref{fig:ratio_graph} profiles the improvement in latency and energy following the optimizations.

\subsection{Impact on Timing and Energy}

Fig.\ref{fig:ratio_graph} profiles the percentage improvement in timing and energy unlocked by \arch{}.

\subsubsection{Common Functional Blocks}

We start with the \texttt{XOR} implementations in Fig.\ref{fig:combxortable}:
design~(a), which comprises three standard gates (NAND, OR, AND), and
design~(b), which uses a NAND and a *OR gate. For the \texttt{XOR} benchmark in
Fig.\ref{fig:ratio_graph}, latency improves by 50\%, directly reflecting an
equivalent reduction in write operations. STT and SOT/SHE achieve energy reductions of 26.0\% and 44.0\%, respectively. State-dependent output resistance renders STT energy sensitive to instruction mix, as the higher serial resistance of Preset-1 instructions elevates operating voltages. Conversely, SOT/SHE maintains consistent resistance and voltage profiles, eliminating this variability and yielding a 91.2\% energy improvement over STT due to lower write current.
%The variability of the STT output resistance is shown in the difference in energy between the NAND and AND instructions, which would otherwise only differ in the direction of their current. This variation is also seen with the energy increase for the *OR operation, as its average energy is raised by instances when the standard voltage is applied to a low output. Preset energy is also a smaller component of preset-1 gates like OR, but is a large fraction of the energy for preset-0 NAND. SOT's uniform operations and significantly reduced write current lead to an order of magnitude better energy cost, and the uniform output resistance leads to more consistent efficiency gain. 

For the multiplexer (\texttt{MUX}) shown in Fig.\ref{fig:ratio_graph}, preset optimization yields a 50\% reduction in both latency and write operations. Energy decreases by 43.1\% for STT and 20.7\% for SOT/SHE. STT benefits from a more favorable instruction mix due to its state-dependent variable resistance. While SOT/SHE exhibits a smaller relative optimization improvement, its inherently lower write current yields a 91.6\% total energy reduction compared to STT. 

Relative to a standard NOR implementation, \texttt{ADD\_NOR} reduces latency by
22.2\% and STT energy by 33.7\%, with negligible impact on SOT/SHE energy.
Transitioning to an unoptimized mixed-gate implementation (\texttt{ADD\_MIX})
reduces latency by 55.5\%, while cutting STT and SOT/SHE energy by 60.5\% and
36.0\%, respectively. Applying preset optimization to the mixed variant
(\texttt{ADD\_OP}) achieves an overall 66.7\% latency reduction, along with STT
and SOT/SHE energy savings of 77.8\% and 63.0\% over the baseline. Comparing {ADD\_MIX} directly to its preset-optimized form reveals an isolated 25\% speedup and a 43.1\% energy improvement.

Finally, the 2-bit comparator (\texttt{COMP}) benchmark achieves a 38.5\% latency reduction, accompanied by STT and SOT/SHE energy reductions of 38.1\% and 45.1\%, respectively.

\subsubsection{Matrix-Vector Multiplication (\texttt{VMUL})}

Multiplication is inherently write-intensive in NVCiM architectures. Preset
optimizations reduce \texttt{VMUL} latency by 15.1\%, which lags the isolated block-level speedup of the full adder due to unoptimized data movement and ancillary overheads. This performance gain stems primarily from reduced instruction latency, supplemented by minor gate count reductions from more efficient \texttt{XOR} logic in half-adders.
STT and SOT/SHE achieve energy reductions of 23.9\% and 13.8\%, respectively. Eliminating inverter (INV) operations (Fig.~\ref{fig:adder_comp}c) yields greater savings for STT because INVs are proportionally more energy-intensive in STT technology. Conversely, SOT/SHE benefits to a lesser degree due to its uniform energy profile and the non-negligible energy cost of data movement outside the adder itself.

\subsubsection{Hopfield Neural Network (\texttt{NN})}

Because a Hopfield network's convergence is data-dependent, performance and
energy metrics are evaluated on a per-iteration basis. As shown in
Fig.\ref{fig:ratio_graph}, optimizing the dot product (\texttt{DOT}) and
selection (\texttt{SEL}) components yields an overall latency reduction of
25.8\%. Speedups achieved at the \texttt{XOR} and \texttt{ADD\_OP} primitive block level are partially
diluted by unoptimized ancillary logic and data movement.
%across the multi-stage pipeline.
%
Preset optimization reduces per-iteration energy consumption by 36.1\% for STT and 18.1\% for SOT/SHE (Fig.\ref{fig:ratio_graph}). The variation between technologies stems from uneven energy scaling across the instruction mix: while SOT/SHE exhibits an order-of-magnitude lower baseline write current, certain operations (such as 3-input AND gates and non-adder instructions in the \texttt{DOT} phase) scale less favorably. Consequently, phase-specific energy savings for the dot product reach 37.6\% for STT and 24.9\% for SOT/SHE.
A minor energy regression occurs in the STT variant of the \texttt{SEL} phase. Here, a bit-scan chain using NAND instructions converts a data-movement step into an inversion, which is energy-expensive due to STT's asymmetric gate energy. Furthermore, because SOT/SHE's absolute write energy is extremely low, low-utilization operations in the \texttt{SEL} phase are dominated by fixed non-gate overheads, dampening relative optimization gains.

\subsection{Impact on Endurance}

%\begin{table}[t]
%    \centering
%    \begin{tabular}{|c|c|c|}
%    \hline
%         & Total Writes & Peak Writes/Cell \\
%         \hline
%        XOR & 6 & 2 \\
%        XOR Opt. & 3 & 3 \\
%        \hline
%        Adder & 8 & 2 \\
%        Adder Opt. & 6 & 4 \\
%        \hline
%        Compare & 26 & 8(2) \\
%        Compare Opt. & 16 & 5\\
%        \hline
%    \end{tabular}
%    \caption{\textbf{Comparisons of Total Writes and Write Density}}
%    \vspace{-.5cm}
%    \label{tab:blockwrites}
%\end{table}

\begin{figure}[t]
    \centering
    \includegraphics[width=1.0\linewidth]{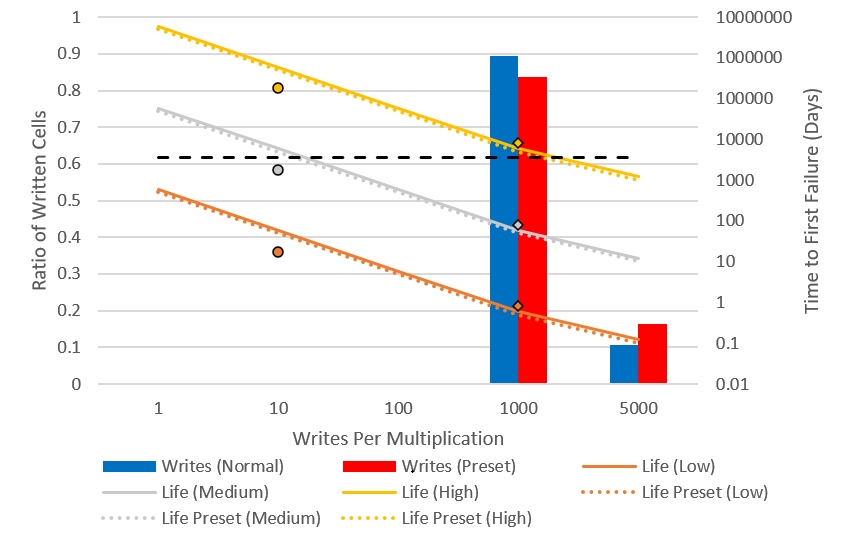}
    \vspace{-.7cm}
    \caption{
    %Distribution of Write Endurance for Vector-Matrix Multiplication Kernel, 
    \texttt{VMUL}'s write endurance.
    %considering three different endurance configurations: {\em Low}, {\em Medium}, and {\em High}.
    Dashed black line marks 10-year time horizon. Diamonds/circles show wear leveling across all outputs/rows.}
    \label{fig:write_dist_vmul}
    \vspace{-.3cm}
\end{figure}

\begin{figure}[t]
    \centering
    \includegraphics[width=1.0\linewidth]{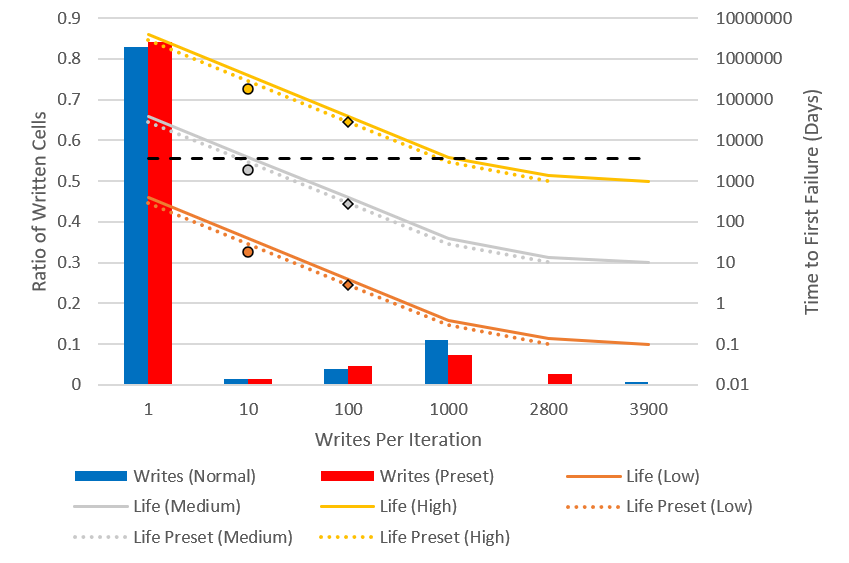}
     \vspace{-.7cm}
    \caption{
    \texttt{NN}'s write endurance.
    %considering three different endurance configurations: {\em Low}, {\em Medium}, and {\em High}.
    Dashed black line marks 10-year time horizon. Diamonds/circles show wear leveling across all outputs/rows in the array.}
    \label{fig:write_dist}
    \vspace{-.3cm}
\end{figure}

\subsubsection{Impact on Functional Blocks}
Preset optimization reduces total write operations by eliminating preset steps
in dependent operations and bypassing both preset and thresholding steps for
elided instructions. As shown in Fig.\ref{fig:ratio_graph}, write reductions
reach 25\% for the optimized mixed-instruction full adder \texttt{ADD\_MIX}, 50\% for
\texttt{XOR}, and 38\% for the magnitude comparator \texttt{COMP}.

However, array longevity also depends on spatial write distributions and cell
endurance limits. Although eliminating redundant operations decreases total
array-wide writes, it often concentrates remaining writes onto fewer destination
cells. This concentrating effect scales with the logical gate depth assigned to
an output cell: for example, the output cell in an optimized \texttt{XOR}
executes a standard NAND followed by a *OR operation, yielding a 50\% write
density increase for that specific cell despite halving total writes. Similarly,
optimized \texttt{MUX} and full-adder blocks exhibit up to a 100\% increase in peak write density.
To counteract localized wear, wear-leveling through operand-address-rotation  can distribute writes across physical memory locations.
At the application level, remapping overhead amortizes over long execution runs~\cite{resch2023endure}.
%At the block level, this involves unrolling code segments or invoking block variants with rotated operand addresses, incurring zero execution latency overhead at the cost of a slightly larger code footprint. At the application level, remapping overhead amortizes over long execution runs; for instance, kernel recompilation every 85,000,000 iterations provides sufficient wear-leveling to mitigate non-uniform write patterns~\cite{resch2023endure}.

%For example, the Full-Adder block has a write footprint of 4 addresses with a uniform 2 writes per cell. The optimized variant has two addresses, one with 2 writes and the other with 4, yielding an average of 3 writes per cell. However, if the optimized block is allowed the same cell footprint and rotates addresses with each invocation, the average writes per cell drops to 1.5. The standard comparator in Fig.\ref{fig:compar_circ} performs 26 writes over 9 cells, averaging 2.89 writes per cell, although avoiding the reuse of temporary variables permits an average of 2 writes per cell. The optimized variant performs 16 writes over 4 cells, which allows wear-leveling in an iso-footprint scenario to reduce the average to 1.78.
%Increasing the pool of cells reduces the average writes per cell linearly. Various methods can be employed, and periodic cycling of addresses at array or device level %used to provide the necessary capacity to reduce the write ratetypically suffices to reduce the write rate \cite{resch2023endure}.

%\subsubsection{Kernel-level Analysis--Work Done versus Lifespan}
\subsubsection{Impact on Work Done versus Lifespan}
\begin{comment}

\begin{table}[t]
    \centering
    \begin{tabular}{|c|c|c|c|}
    \hline
         & TTFF (Low) & TTFF (Medium)  & TTFF (High)  \\
         \hline
        Default & 0.12  & 12.17 &	1217.07 \\
        \hline
        Preset/Delay & 0.10	& 10.33&	1032.87  \\
        \hline
        Preset Only  & 0.12 & 11.87 &	1187.43  \\
        \hline
    \end{tabular}
    \caption{Vector Matrix Multiply Time to First Failure (days) Versus Maximum Cell Writes. }
    \label{tab:Vmulendurance}
\end{table}

\begin{table}[t]
    \centering
    \begin{tabular}{|c|c|c|c|}
    \hline
         & TTFF (Low) & TTFF (Medium)  & TTFF (High)  \\
         \hline
        Default & 0.10 & 9.98 &  998.03 \\
        \hline
        Preset/Delay & 0.10 & 10.3 & 1029.90  \\
        \hline
        Preset Only  & 0.14 & 13.89 & 1389.29  \\
        \hline
    \end{tabular}
    \caption{Neural Network Time to First Failure (days) Versus Maximum Cell Writes. }
    \label{tab:NNendurance}
    %\vspace{-.2cm}
\end{table}

\end{comment}

\begin{table}[t]
    \centering
    \begin{tabular}{|c|c|c|}
    \hline
         &  \texttt{VMUL}  & \texttt{NN}  \\
         \hline
        {\em Default} & 1217.1 & 1000.31 \\
        \hline
        {\em Preset/Delay} &	1032.9 & 1014.9 \\
        \hline
        {\em Preset Only}  &	1187.4 & 1228.4 \\
        \hline
    \end{tabular}
    \vspace{.1cm}
    \caption{Time to First Failure---TTFF (days) at {\em High} endurance (without wear leveling). }
    \label{tab:VmulNNendurance}
    \vspace{-.7cm}
\end{table}

The lower dashed lines in Figs.~\ref{fig:write_dist_vmul}
and~\ref{fig:write_dist} highlight a primary tradeoff of preset optimization:
while removing instructions reduces execution time and energy, it does not
  necessarily decrease write intensity on the peak-write cells that bound
  endurance. Consequently, wall-clock endurance decreases across almost all
  preset-optimized configurations. In \texttt{VMUL}, execution time scales
  directly with instruction count; thus, removing presets accelerates
  computation as much as it cuts total writes. As summarized in
  Table~\ref{tab:VmulNNendurance}, time-to-first-failure (TTFF) regresses in the
  \arch{}-optimized case because eliding instructions reduces total runtime
  without effectively relieving write stress on the bottleneck (peak-write) cells.

Table~\ref{tab:VmulNNendurance} details the endurance impact of preset optimization across equivalent computations. Lacking targeted wear-leveling techniques~\cite{resch2023endure}, the baseline configuration exhibits a short time-to-first-failure (TTFF), lasting approximately $2.4~\text{hours}$ under conservative limits. The \textit{Preset/Delay} configuration, which accelerates instruction execution, yields minimal wall-clock endurance gains due to increased operational throughput. Conversely, the \textit{Preset Only} configuration retains standard instruction latency, translating preset optimizations directly into a $22.8\%$ operational lifespan extension.
However, evaluating wall-clock lifespan in isolation overlooks the throughput efficiency of \textit{Preset/Delay}, which completes each unit of work in $0.74\times$ the baseline time. For domain-specific accelerators, total work completed over the device lifetime provides a more representative metric than raw operating hours alone.

\subsubsection{Impact on Wear Leveling}
Fig.\ref{fig:write_dist_vmul} and Fig.\ref{fig:write_dist} present a holistic
evaluation of array lifespan by clustering memory cells according to write
activity. Specifically, we consider two wear-leveling
strategies:
%~\cite{resch2023endure}\red{
\textit{active-only rotation}, which
distributes logical addresses exclusively among cells in the active
computational footprint, and \textit{array-wide rotation}, which shifts write
operations across all available physical rows in a $1024$-row array. 
%[please double-check]} 
Active cells experience highly disparate write levels:
\texttt{VMUL} exhibits a tight distribution spanning an order of magnitude,
whereas the more complex \texttt{NN} benchmark displays a much broader spread. Peak write categories capture temporary scratch cells within the inner computational loops, whereas inputs and neural weights in \texttt{NN} undergo massive reuse across iterations, resulting in negligible write traffic.

Wear-leveling effectively mitigates this write concentration. Across both
kernels, active writes occur in only several dozen cells per column, with just 5
to 7 cells bearing peak traffic. In \texttt{NN}, $83\%$ of the data footprint is
rarely updated, while a small fraction absorbs $10\times$ to $1000\times$ the
write volume. Although preset optimizations alone cannot bridge
orders-of-magnitude write imbalances, they reduce the fraction of highest-demand
cells from $11.1\%$ to $9.86\%$. This $15.3\%$ reduction in peak-cell count
yields a disproportionate endurance benefit because these bottleneck cells
dominate failure rates. As shown by the diamond markers in
Fig.\ref{fig:write_dist}, active-only wear-leveling across the 154 active cells
reduces writes per iteration from 139 (baseline) to 106 (optimized), extending
lifespan beyond $10~\text{years}$ for the \textit{High} endurance
configuration. Full array-wide rotation across all 1024 rows yields over
$5~\text{years}$ of lifespan for the \textit{Medium} endurance configuration,
whereas the \textit{Low} endurance case requires multi-array wear-leveling.

Because \texttt{VMUL} performs a single dedicated function, it heavily utilizes
a small set of temporary variables and accumulators. The preset-optimized
\texttt{VMUL} kernel reduces execution latency, which increases the average write rate while concentrating remaining writes onto fewer cells. Consequently, Fig.\ref{fig:write_dist_vmul} shows that active-only wear-leveling provides smaller relative gains for \texttt{VMUL} due to its smaller active footprint, though array-wide rotation yields endurance improvements comparable to \texttt{NN}.

\section{{Related Work}}
\label{sec:relwork}
%\subsection{Related Work}
Output cell reuse for specific logic operations has been explored in ReRAM-based architectures (e.g., SCRIMP~\cite{scrimp20} for XNOR and FELIX~\cite{felix2018} for XOR) as well as MRAM-based designs such as BSCIM~\cite{wangsotserial}. Architectures like FELIX inverse-map resistance levels to logical states ($0/1 \rightarrow \text{high/low}$)~\cite{felix2018,magic,xmagic}, which corresponds to inverting the Preset-0 and Preset-1 {categories in Table \ref{tab:presetcategory}}. Similarly, X-MAGIC~\cite{xmagic} and ILOSS~\cite{iloss} analyze specific, destructive gate combinations to conserve output cells. Crucially, while these prior works are restricted to a few architecture-specific primitives, \arch\ provides a generalized optimization framework. For example, whereas X-MAGIC's uninitialized NOT instruction implements $A\overline{B}$~\cite{xmagic}, \arch's MRAM-based \texttt{*INV} implements $A + \overline{B}$---demonstrating that \arch's generalized methodology encompasses diverse architectural mappings and subsumes prior targeted techniques.

\section{{Conclusion}}
However, fully realizing NVCiM's potential requires addressing a fundamental
operational inefficiency: conventional digital NVCiM designs require each target
output cell to undergo an operation-specific ``preset'' write to establish a deterministic state prior to logic evaluation. In multi-gate computational cascades, these redundant preset writes dominate energy consumption and accelerate structural cell degradation, frequently neutralizing CiM's intrinsic efficiency advantages.

To resolve this bottleneck, we present \textbf{PRIM} (\textbf{P}reset
\textbf{R}eduction \textbf{I}n \textbf{M}emory), a specialized compiler
optimization framework for digital NVCiM. PRIM tracks logic dependencies and
intermediate state transitions across arbitrary Boolean networks to
systematically bypass redundant preset and thresholding operations. Crucially,
PRIM requires zero hardware or architectural modifications and generalizes across arbitrary logic sequences rather than relying on hardcoded gate primitives.

Although \arch{}-induced write reductions lower total array traffic, they do not always yield a proportional decrease in per-cell write intensity—a critical factor governing overall array lifetime. In fact, \arch{} can concentrate remaining operations, increasing the localized write intensity on specific output cells. We quantitatively analyze this tradeoff and demonstrate that the benefits of \arch{} optimization persist when combined with standard wear-leveling techniques. Integrating \arch{} with existing wear-leveling schemes is straightforward, as \arch{} is non-invasive and broadly applicable across diverse digital NVCiM architectures.

Evaluated across diverse computational kernels and NVCiM technologies, we
demonstrate that PRIM achieves up to a 50\% reduction in execution latency, up to a 44\% reduction in energy consumption, and up to a 50\% improvement in cell endurance. By eliminating redundant write cycles without compromising computational accuracy, PRIM establishes a practical path toward scalable, ultra-low-energy digital NVCiM systems.

\section{{Appendix: Common Computational Primitives}}
\label{sec:app}
%\red{1. Please make sure that the format, terminology, ... is consistent with main text. For benchmarks, we now use ``texttt'', as in \texttt{ADD\_MIX}. We can keep the gate-level representation as is here otherwise---please check e-mail comments on this.}
%\blue{Cleaned up diagrams in appendix. }

%\red{2. Would be good to add a little bit more insight and a brief explanation of for Algorithm~\ref{alg:blockopt} here: what the masks correspond to etc., making sure that all parameters/variables are clearly defined.}
%\blue{added description of algorithm at end of appendix, should this move somewhere else?}
%\red{2. Please make sure that Fig.~\ref{fig:tree} is properly referenced here, with a brief explanation.}

\begin{algorithm}[t]
%\caption{Exhaustive Block Optimization}
%\caption{End-to-End \arch\ Optimization Example}
\caption{End-to-end \arch\ Optimization}
%to enforce the Preset-0 and Preset-1 rules}
\footnotesize
\begin{algorithmic}[1]
\State Given N-bit input region, generate bit-mask size $2^N$ (maskT) based on truth table.
\State Initialize node with list of unused instruction and input combinations. Initialize maskA of asserted values status to 0.
\ForAll {Combinations} 
%\State Select a gate from set of unused input and gate combinations.
\State Set maskI using instruction's truth table over inputs.
\If{$preset == 0$}
\ForAll{$maskI_k$} 
\State temporary $maskA_k'=maskI_k$ . 
\EndFor
\Else
\ForAll{$maskI_k$} 
\State temporary $maskA_k'=inverse(maskI_k)$ . 
\EndFor
\EndIf
\If {no $maskA_k' == 1$ when $maskA_k==0$}
\State signal $Reject$
\EndIf
\ForAll {$maskA_k == 0$}
\If {$maskA' == 1$ and $maskI\neq maskT$}
\State signal $Reject$  
\EndIf
\EndFor
\If {$Reject$}
\State Continue to next combination.
\Else
\State Accept instruction. 
\State $maskA =$ bitwise OR($maskA$,$maskA'$) 
\If {$maskA$ is all 1}
\State mark path complete
\EndIf
\State Recursively call initialization for child nodes, override new maskA with current maskA.
\State Optionally set max tree depth based on gate count of standard circuit.
\EndIf
\EndFor
%If the result exists in the stack, reject. Otherwise update Pmask, use mask and return to selection.
%\State If rejecting, return to select.
%\State Repeat until Pmask matches the inverse truth table of an instruction, and use standard version of that operation as final .
%\State If successful, the sequence of instructions is a valid preset-optimized chain, in reverse order. 
%\State A dead end is a candidate for storing the current result, and using another cell for the remaining mask bits. Combine cells with standard ops.
\end{algorithmic}
\label{alg:blockopt}
\end{algorithm}

\begin{figure}[t]
    \centering
    \includegraphics[width=1.0\linewidth]{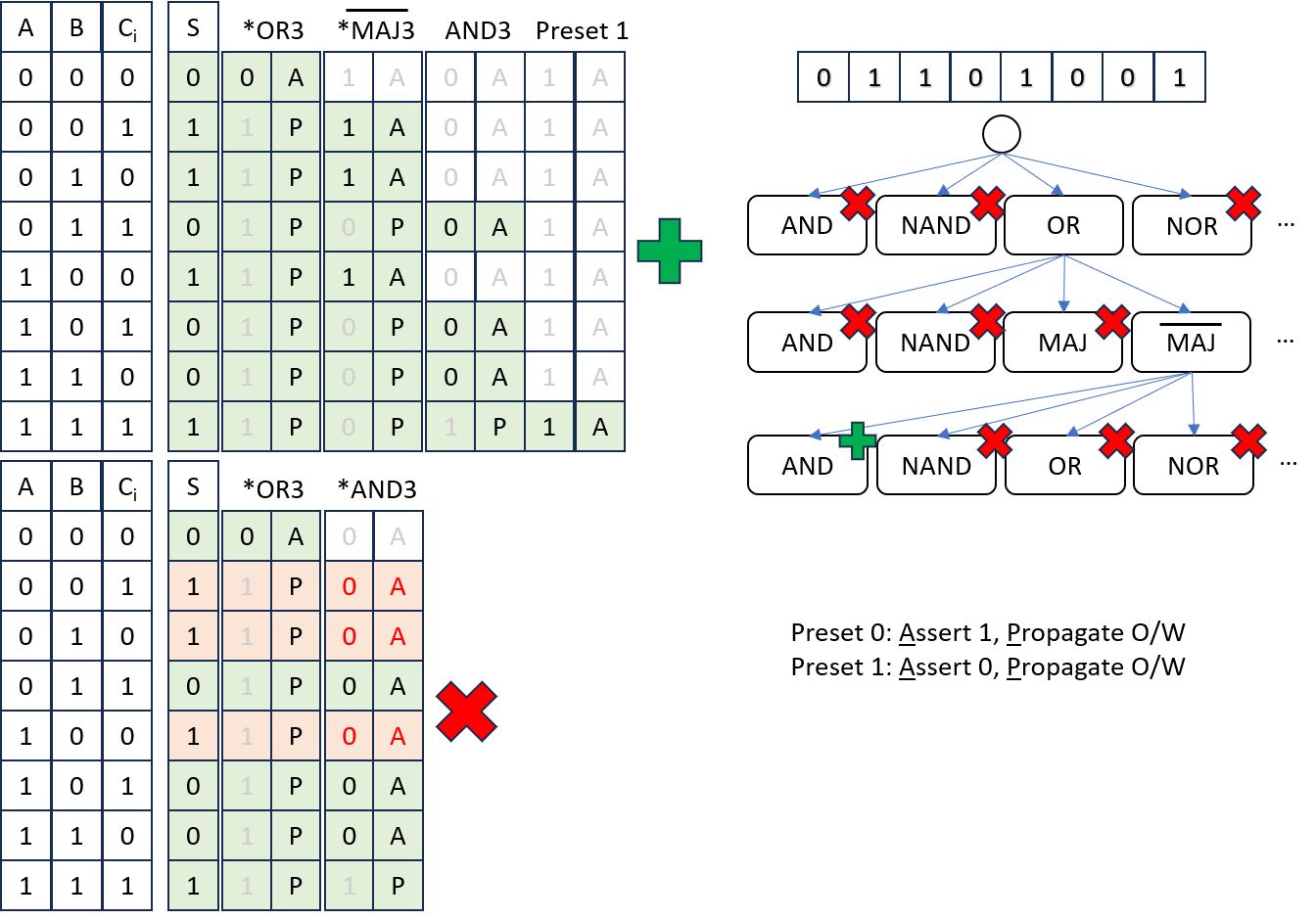}
    \vspace{-0.2cm}
    \caption{Generation and traversal of state transition tree for valid and invalid preset optimizations for full-adder sum.}
    \label{fig:tree}
    \vspace{-.2cm}
\end{figure}

%red{needs a consistency pass (wrt front-end, including fig formats)}
For many computational kernels, the instruction stream is dominated by the repeated invocation of a common sub-blocks. For example, division, multiplication, and multiply-accumulate (MAC) in matrix/vector math are composed streams of full-adders. %The latter type of workload can be bandwidth-bound, making CiM compelling.

\begin{figure}[tp]
%\centerline{\includegraphics[scale=0.25]{figures/adder3.png}}
\centerline{\includegraphics[scale=0.475]{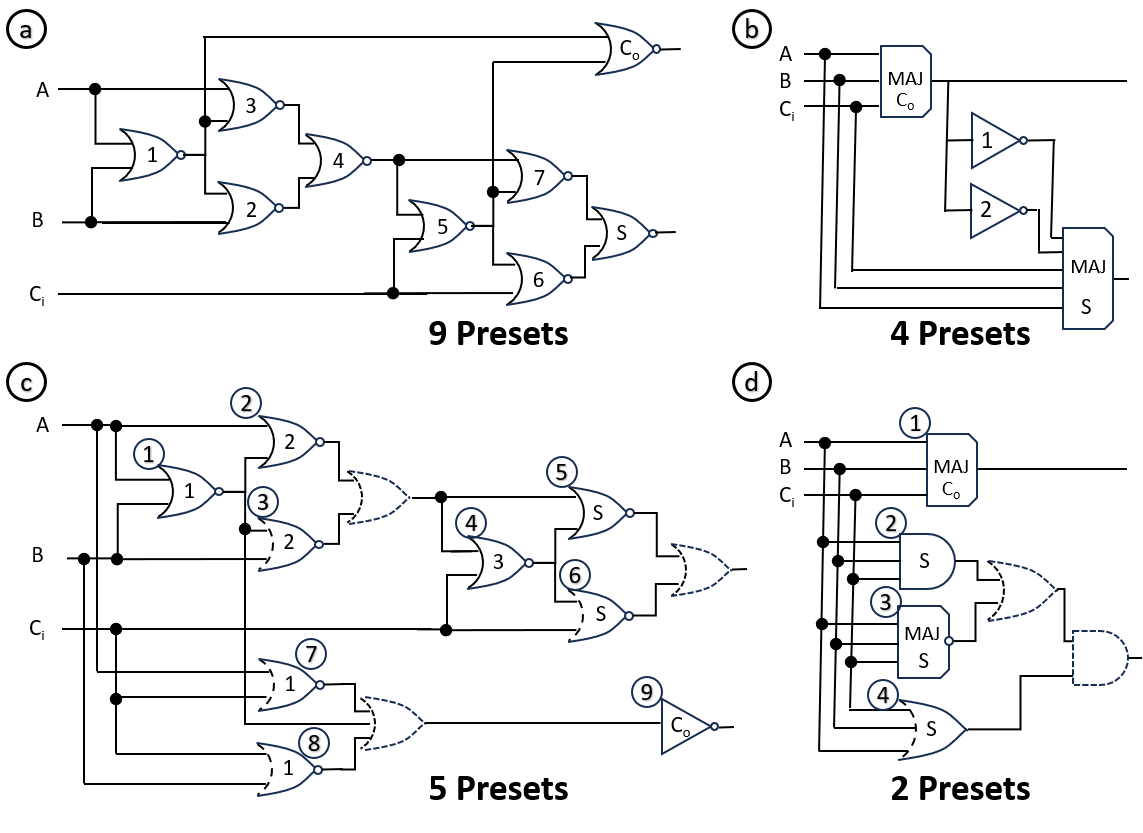}}
\vspace{-0.1cm}
\caption{
Full-adder implementations: a) NOR gate, b) Mixed gate, c) NOR with \arch, d) Mixed gate with \arch. For compactness, array activity is denoted by equivalent gate symbols. Inside each symbol is the associated temp or output variable (majority gates also labeled with "MAJ"). In PRIM circuits, numbered gates with dashed inner arcs have no preset, blank fully-dashed outlines represent the logical effects of \arch, and numbers in circles show order of operation. 
}
\vspace{-0.5cm}
\label{fig:adder_comp}
\end{figure}

\noindent {\bf Full-Adder:} 
%Algorithm \ref{alg:FA} provides the NVCiM implementation of a {Full-Adder} as a common functional block. This implementation featuring majority (MAJ) gates is optimized to minimize the gate count (assuming atomic preset and thresholding), which is the default optimization goal during application mapping. MAJ gates detect when the majority of their inputs are high to produce the outputs for the carry and sum. 
%
%Algorithm 8 provides the optimized version. Here the calculation of the Sum bit avoids presets using a $\overline{\mbox{MAJ3}}$ instruction which returns 1 if the majority of its three inputs are 0.
%, and optimizations have been made to implement them in as few gates as possible. Algorithm \ref{alg:FA} uses 
%While the operation count remains the same, the *$\overline{\mbox{MAJ3}}$ and *OR gates reduce the write count and could be merged into one instruction. The standard algorithm can be run with a total of three instructions if the array supports a 2-output INV gate, or four otherwise\cite{cram}.
%
%A full-adder is a logical block that serves as a component of many computational primitives, such as many forms of multi-bit adder, as well as multipliers and dividers. Matrix and vector arithmetic are memory-intensive workloads that can benefit from the internal bandwidth of a CiM, and can be implemented as deep chains of full-adder invocations.
%
Fig.\ref{fig:adder_comp} outlines a continuum of full-adder gate implementations. Fig.\ref{fig:adder_comp}a shows an adder composed of 9 standard NOR gates, which performs 18 writes. %The input cells are represented as green cells under the diagram, and cells used as the outputs for the respective gates, each receiving one preset and one preset step. 
%This results in a total of 18 writes for the full-adder.

Fig.\ref{fig:adder_comp}b (\texttt{ADD\_MIX}) shows the impact of an NVCiM that supports a range of operations, like majority functions with 3 or 5 inputs (MAJ3,MAJ5), and inversion (INV) gates. Using this varied mix can bring the cell count to 4 and 8 writes.

Fig.\ref{fig:adder_comp}c (\texttt{ADD\_NOR}) shows a NOR-based full-adder utilizing \arch. This compresses the number of output cells to 5, although several must be used repeatedly. This requires 14 writes. 

Fig.\ref{fig:adder_comp}d (\texttt{ADD\_OP}) applies \arch ~to \texttt{ADD\_MIX}, rewriting the sequence to use $C_{out}=MAJ3$, $S=AND3\xrightarrow{}\overline{*MAJ3}\xrightarrow{}*OR3$.  The footprint shrinks to the minimum possible 2 outputs, while reducing the writes to 6.
%\begin{algorithm}[t]
%\caption{Full-Adder \cite{cram}}
%\small
%\begin{algorithmic}[1]    
%    \State $C_{out}  = MAJ3(A,B,C_{in})$
%    \State $tmp_1 = INV(C_{out})$
%    \State $tmp_2 = INV(C_{out})$
%    \State $Sum = MAJ5(A,B,C_{in},tmp_1,tmp_2)$
%    \end{algorithmic}
%    \label{alg:FA}
%\end{algorithm}

\begin{figure}[t]
%\centerline{\includegraphics[scale=0.3]{figures/mux3.png}}
\centerline{\includegraphics[scale=0.45]{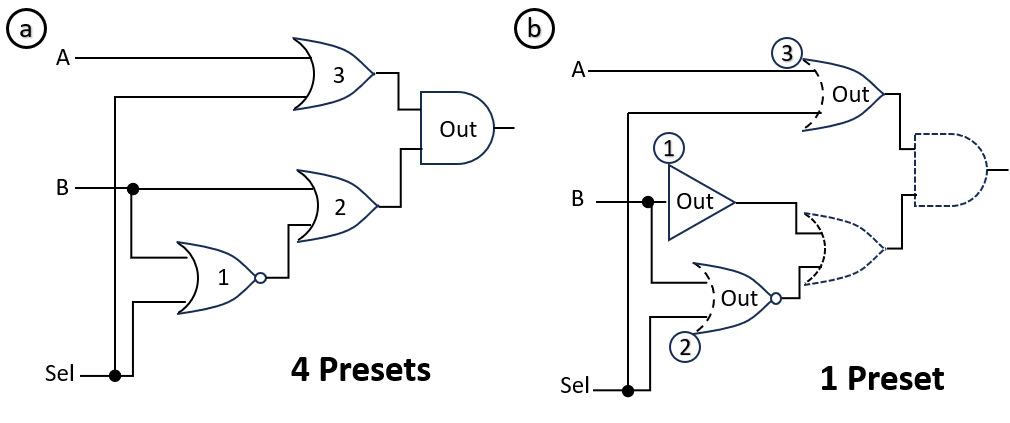}}
\vspace{-0.1cm}
\caption{
2:1 Multiplexer: a) Standard gate, b) Using \arch. 
%Numbered gates with dashed inner arcs have no preset, blank fully-
%dashed outlines represent effect of \arch. Numbers in circles show order of operations.
}
\vspace{-0.3cm}
\label{fig:muxdiag}
\end{figure}

\noindent{\bf Multiplexer:} Fig.\ref{fig:muxdiag} covers the implementation of a {multiplexer (\texttt{MUX})}, another common block. It can be used in computational kernels to implement variable shifts or selections using dynamic values.
%{\bf Multiplexers} are a simple functional block, which can be 
%MUXes used in cases where an algorithm must select between inputs. 

As shown by Fig.\ref{fig:muxdiag}a, the example \texttt{MUX} output (Out) takes input from A or B, based on selection variable Sel. A 2:1 \texttt{MUX} can be scaled to larger input counts by combining multiple instances in a multiplexer tree: such as a 4:1 \texttt{MUX} made of a pair of 2:1 MUXes that feed a final 2:1 \texttt{MUX}. 

The optimized Fig.\ref{fig:muxdiag}b benefits from multiple inputs that feed into a single output, and can be done with only one cell. Often, multiple sequences can match a truth table. The chosen sequence directly maps gates in Fig.\ref{fig:muxdiag}a to implied gates in Fig.\ref{fig:muxdiag}b.

%\begin{enumerate}
%    \item $C_{out}=MAJ3(A,B,C_{in})$
%    \item $Sum = AND(A,B,C_{in})$
%    \item $Sum = *\overline{MAJ3}(A,B,C_{in})$
%    \item $Sum = *OR(A,B,C_{in})$
%\end{enumerate}
%
%\begin{algorithm}[t]
%\caption{Full-Adder (Optimized)}
%\small
%\begin{algorithmic}[1]    
%    \State $C_{out} = MAJ3(A,B,C_{in})$
%    \State $Sum = AND(A,B,C_{in})$
%    \State $Sum = *\overline{MAJ3}(A,B,C_{in})$
%    \State $Sum = *OR(A,B,C_{in})$
%    \end{algorithmic}
%    \label{alg:FA2}
%\end{algorithm}

%When evaluating the truth table for the Sum value, all but two of the entries in the table are covered by the $\overline{\mbox{MAJ3}}$ gate, with the exceptions being $A=B=C_{in}=1$ handled by the AND, and an $A=B=C_{in}=0$ handled by the OR. 

\begin{comment}
\begin{algorithm}[t]
\small
    \caption{2:1 Multiplexer}
    \begin{algorithmic}[1]
    \State $tmp_1 = NOR(B,Sel)$
    \State $tmp_2 = OR(tmp_1,B)$
    \State $tmp_3=OR(A,Sel)$
    \State $Out = AND(tmp_2,tmp_3)$
    \end{algorithmic}
    \label{alg:mux}
\end{algorithm}

\begin{algorithm}[t]
\small
    \caption{2:1 Multiplexer (Optimized)}
    \begin{algorithmic}[1]
    \State $Out = COPY(B)$
    \State $Out = *NOR(B,Sel)$
    \State $Out = *OR(A,Sel)$
    \end{algorithmic}
    \label{alg:muxopt}
\end{algorithm}
\end{comment}

%Per Table \ref{tab:energies}, endurance impact is improved by 50\%, but with the caveat that this has concentrated the remaining writes on one output cell. Some form of wear-leveling or periodic shifting of destination address would be needed to reduce the localized impact. STT and SOT demonstrate energy improvements of 26.0\% and 44.0\%, respectively. The variability in output resistances leads to less even improvement for the STT energy, while SOT's consistent resistances and voltages allow for better improvement on top of the large improvement in write energy it has over STT.

\noindent {\bf Comparator:} Comparators are also 
%are another example of potentially 
useful functional blocks. The SIMD structure of the NVCiM relies on evaluating conditional paths in a branchless manner, where outcomes are masked by test flags. Magnitude comparators provide flags for less-than, greater-than, and equality. 

The comparator shows the benefits and limits of optimizing presets.
%and do not map to the CiM substrate as cleanly as the full adder.
A 1-bit magnitude comparator takes bits A and B, and provides three outputs regarding the relationship of A to B, less than ({\em LT}), greater than({\em GT}), and equal ({\em EQ}), respectively. 
%\begin{enumerate}
%    \item $temp=NAND(A,B)$
%    \item $LT = AND(temp,B)$
%    \item $GT = AND(temp,A)$
%    \item $EQ = NOR(LT,GT)$
%\end{enumerate}
%\begin{algorithm}[t]
%\small
%    \caption{1-bit Magnitude Comparator}
%    \begin{algorithmic}[1]
%    \State $tmp = NAND(A,B)$
%    \State $LT = AND(tmp,B)$
%    \State $GT = AND(tmp,A)$
%    \State $EQ = NOR(LT,GT)$
%    \end{algorithmic}
%    \label{alg:comp1}
%\end{algorithm}

%The simple truth table and requirement of three separate outputs limits the preset optimization options.
%\begin{enumerate}
%    \item $LT=INV(A)$
%    \item $LT = *COPY(B)$
%    \item $GT = INV(B)$
%    \item $GT = *COPY(A)$
%    \item $EQ = NOR(LT,GT)$
%\end{enumerate}

%\begin{algorithm}[t]
%\caption{1-bit Magnitude Comparator (Optimized)}
%\small
%\begin{algorithmic}[1]
%    \State $LT = INV(A)$
%    \State $LT = *COPY(B)$
%    \State $GT = INV(B)$
%    \State $GT = *COPY(A)$
%    \State $EQ = NOR(LT,GT)$
%    \end{algorithmic}
%    \label{alg:comp2}
%\end{algorithm}
The larger number of output cells than inputs causes PRIM to be counterproductive, as unique outputs require at least one preset to initialize, and the block lacks a depth of dependent gates.
%Algorithms \ref{alg:comp1} and \ref{alg:comp2} show that in this case, multiple outputs and the limited dependence chain lead to an expansion. A non-standard gate such as a two-output NAND would be needed to replace the two INV instructions, but would only remove one write overall.
%\blue{comparator is more recent and larger, needs more complete validation for the versions correctness}
A two-bit comparator (\texttt{COMP}) provides the same outputs while comparing two-bit numbers. 
%An example implementation with standard gate instructions can supply EQ in 7 gates, LT in 5, and GT is the NOR of the previous two. As each gate is a two-step operation, this translates to 26 writes overall.
%\begin{enumerate}
%    \item $t1=AND(A_0,B_0)$
%    \item $t2=NOR(A_0,B_0)$
%    \item $t3 = OR(t1,t2$
%    \item $t1=AND(A_1,B_1)$
%    \item $t2=NOR(A_1,B_1)$
%    \item $t4 = OR(t1,t2)$
%    \item $EQ = AND(t3,t4)$
%    \item $t1=NAND(A_0,A_1)$
%    \item $t2=OR(A_1,B_1)$
%    \item $t3=AND(t1,t2,B_0)$
%    \item $t1=NOR(A_0,B_0,EQ)$
%    \item $LT = OR(t1,t3)$
%    \item $GT = NOR(LT,EQ)$
%\end{enumerate}
\begin{figure}[t]
    \centering
    \includegraphics[width=0.8\linewidth]{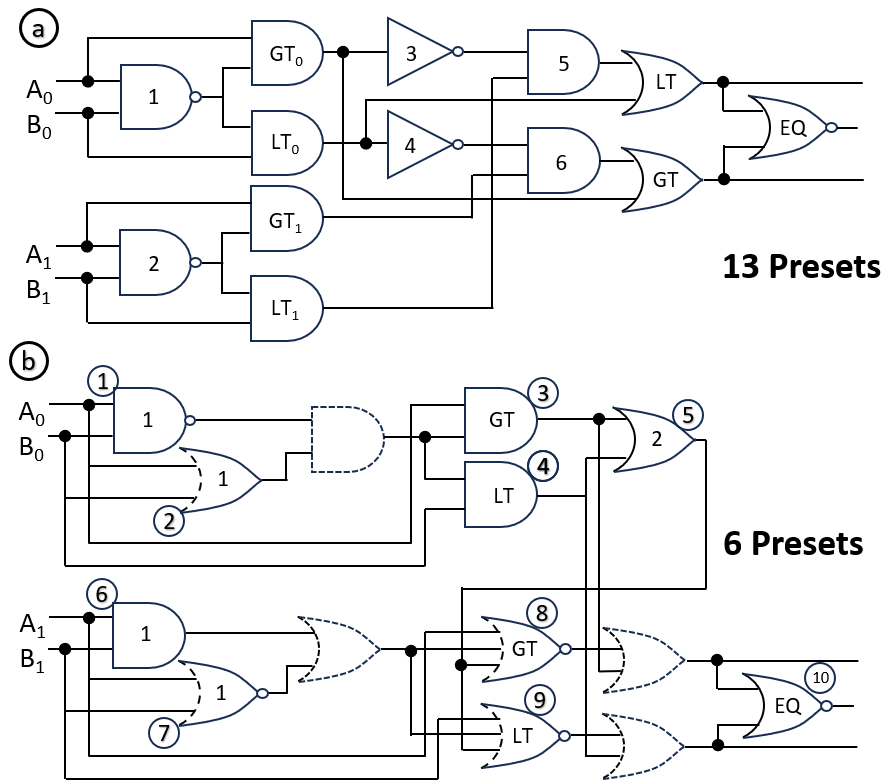}
    \caption{2-bit Magnitude Comparator: a) Non-optimized b) Optimized variant. %Gates with dashed borders are implicit operations. Labeled squares represent cell usage. 
    }
    \label{fig:compar_circ}
    \vspace{-.3cm}
\end{figure}
Fig.\ref{fig:compar_circ}a cascades two standard 1-bit comparators and combines their results in 13 instructions and 26 writes.
%\begin{enumerate}
%    \item $temp=NAND(A_0,B_0)$ stage 1
%    \item $GT_{prev}=AND(temp,A_0)$
%    \item $LT_{prev}=AND(temp,B_0)$
%    \item $temp=NAND(A_1,B_1)$ stage 2
%    \item $GT_1=AND(temp,A_1)$
%    \item $LT_1=AND(temp,B_1)$ 
%    \item $temp=INV(LT_{prev})$ combine results
%    \item $temp_1=AND(temp,GT_1)$
%    \item $GT=OR(GT_0,temp_1)$  $//GT=GT_0+\overline{LT_{prev}}GT_1$ 
%    \item $temp=INV(GT_{prev})$
%    \item $temp_1=AND(temp,LT_1)$
%    \item $LT=OR(LT_{prev},temp_1)$ $//LT=LT_0+\overline{GT_{prev}}LT_1$ 
%\item $EQ=NOR(GT,LT)$ final NOR
%\end{enumerate}
\begin{comment}
    
\begin{algorithm}[t]
\caption{2-bit Magnitude Comparator}
\small
\begin{algorithmic}[1]
    \State $tmp_1 = NAND(A_0,B_0)$ stage 1
    \State $GT_{prev} = AND(tmp_1,A_0)$
    \State $LT_{prev} = AND(tmp_1,B_0)$
    \State $tmp_1 = NAND(A_1,B_1)$ stage 2
    \State $GT_1 = AND(tmp_1,A_1)$
    \State $LT_1 = AND(tmp_1,B_1)$ 
    \State $tmp_1 = INV(LT_{prev})$ combine results
    \State $tmp_2 = AND(tmp_1,GT_1)$
    \State $GT = OR(GT_0,tmp_2)$  $//GT=GT_0+\overline{LT_{prev}}GT_1$ 
    \State $tmp_1 = INV(GT_{prev})$
    \State $tmp_2 = AND(tmp_1,LT_1)$
    \State $LT = OR(LT_{prev},tmp_2)$ $//LT=LT_0+\overline{GT_{prev}}LT_1$ 
    \State $EQ=NOR(GT,LT)$ 
    \end{algorithmic}
    \label{alg:comp_casc}
\end{algorithm}
\end{comment}
\texttt{COMP} can cascade to higher bit counts by repeating the 6th and later gates, with the prior stage's {\em LT} and {\em GT} becoming the $GT_{0}$ and $LT_{0}$ in the next. 
%It can be observed 
We observe that storing stage results into the flag variables requires 2$\times$ the gates than are in the stage itself: an example of the $C=AB+C$ issue.

Fig.\ref{fig:compar_circ}b shows that presets can be removed to optimize cascaded logic. The {\em GT} and {\em LT} flags are ``sticky'' bits that, once one is set exclusive of the other, stop future changes. The *NOR gates 8-9 link stages. %The EQ flag does not need to be calculated until after the final bit is processed.
%\begin{enumerate}
%\item $temp=NAND(A_0,B_0)$ stage 1
%    \item $GT=AND(temp,A_0)$
%    \item $LT=AND(temp,B_0)$
%    \item ...
%    \item *stage k*
%    \item $EQ_k=NOR(GT_k,LT_k)$
%    \item $GT=*NOR(LT,LT_k,EQ_k$)
%    \item $LT=*NOR(GT,GT_k,EQ_k$)
%    \item ...
%    \item $EQ=NOR(GT,LT)$ final NOR
%\end{enumerate}
\begin{comment}
\begin{algorithm}[t]
\caption{$k^{th}$-Stage Cascading Logic for Comparator}
    \small
    \begin{algorithmic}[1]
    \State $EQ_k = NOR(GT_k,LT_k)$
    \State $GT = *NOR(LT,LT_k,EQ_k$)
    \State $LT = *NOR(GT,GT_k,EQ_k$)
    \end{algorithmic}
    \label{alg:compstage}
\end{algorithm}
\end{comment}
%The *NOR operations without a preset will ensure the flags remain at 1 once set, as the non-preset component of a NOR operation cannot pull down the current value.
%This reduces a 2-bit comparator to 10 instructions and 18 writes.
While preset optimizations fare poorly on a 1-bit comparator, Fig.\ref{fig:compar_circ}b shows a 2-bit version stores state more efficiently.
%\begin{enumerate}
%    \item $t1=AND(A_0,B_0)$
%    \item $t1=*NOR(A_0,B_0)$
%    \item $EQ=AND(A_1,B_1)$
%    \item $EQ=*NOR(A_1,B_1)$
%    \item $EQ = *COPY(t1)$ 
%    \item $LT=NAND(A_0,A_1)$
%    \item $LT=*AND(B_0,B_1)$
%    \item $LT=*NOR(A_0,EQ)$
%    \item $LT = *OR(B_0,B_1)$
%    \item $GT = NOR(LT,EQ)$
%\end{enumerate}
\begin{comment}
\begin{algorithm}[t]
\caption{2-bit Comparator (Optimized)}
    \small
    \begin{algorithmic}[1]
    \State $tmp_1 = AND(A_0,B_0)$
    \State $tmp_1 = *NOR(A_0,B_0)$
    \State $EQ = AND(A_1,B_1)$
    \State $EQ = *NOR(A_1,B_1)$
    \State $EQ = *COPY(tmp_1)$ 
    \State $LT = NAND(A_0,A_1)$
    \State $LT = *AND(B_0,B_1)$
    \State $LT = *NOR(A_0,EQ)$
    \State $LT = *OR(B_0,B_1)$
    \State $GT = NOR(LT,EQ)$
    \end{algorithmic}
    \label{alg:compop}
\end{algorithm}
\end{comment}
Unlike simpler blocks, evaluation involves extra cells besides the three outputs. %The XOR and XNOR operations and an OR gate tracking whether sticky bits have been set need separate cells.
The function has states that cannot be distinguished using a shared cell, and a separate OR allows for the flags' current states to control future updates.
%\begin{enumerate}
%    \item stage body
%    \item $GT=*NOR(LT,LT_k,EQ_k)$ combine
%    \item $LT=*NOR(GT,GT_K,EQ_k)$
%    \item ...
%    \item $EQ=NOR(GT,LT)$ final NOR
%\end{enumerate}
Maintaining the flag values incurs an additional two gates with two writes, and a final standard NOR to compute the final {\em EQ} value. Cascading to higher bit counts can be accomplished by repeating the 5th gate and later. The 2-bit variant uses 10 instructions and requires 16 writes.

\noindent {\bf Additional Blocks:} Optimizing other blocks is a boolean optimization problem, where the rules for PRIM can weight optimization algorithms towards gate sequences that fit the cost model unique to an NVCiM. Algorithm \ref{alg:blockopt} exhaustively searches for optimal chains of operations that can be mapped to a single output cell, if a shorter than standard chain exists. If PRIM cannot provide a total solution, multiple sequences can be mapped to separate cells and connected by gate operations. 

The algorithm works backwards from the bit-mask $maskT$, which is populated with the binary output of the target function across all entries of its truth table. The algorithm builds a level of a tree structure, with the potential nodes corresponding to each combination of gate operation and input operands. Preset-0/1 instructions Assert 1/0 if the threshold is met, and propagate 0/1. A propagation mask $maskA$ tracks which entries in $maskT$ have been asserted by a gate operation, and a local $maskA'$ based on the current gate's truth table is compared to the in-progress mapping. If the candidate gate can assert any unresolved mask entries without conflicting with the target value or contradicting what is already set, the instruction is accepted and the local $maskA$ is updated. The algorithm can recursively generate a subtree below the current node, until success or failure. Once done, the algorithm moves back up the levels of the tree and evaluates all candidate gate/operand options at each level. A breadth-first search of the tree, as seen in Fig. \ref{fig:tree} will find the shortest path in reverse order.

Fig.\ref{fig:tree} shows a visual representation of the process applied to the $Sum$ output of \texttt{ADD\_OP}. A valid selection of *OR3 lists its truth table as well as a list of entries it asserts (A) or propagates (P). Subsequent operations compare their truth tables and asserted values, until eventually all of the target truth table's entries are asserted correctly. An incorrect entry would assert a value that conflicts with the target function, shown as red rows in the bottom table. $maskA'$ indicates some values in $maskI$  conflicts with $maskT$, and only propagation (P) flags exist between the current node and the root.

This exhaustive search for the most optimum solutions can experience a combinatorial explosion of gates and input addresses. Growth is constrained by the max depth set by a standard sequence, limited gate types, and fan-in. Table \ref{tab:presetcategory} lists more gate types than many architectures have, and fan-in is noise-limited  \cite{zabihi2019spinhall,zabihi2020analyzing}. Diminishing returns and computational cost generally cap the footprint for PRIM. Expensive evaluations built into a compiler library can amortize the cost.

%\blue{Overlaps with description in page 6. Remove some of the details here, or streamline in page 6?}
%\red{we can leave the front-end as is, and prune here a bit to fit into 11 pages.}

%%%%%%% -- PAPER CONTENT ENDS -- %%%%%%%%

%%%%%%%%% -- BIB STYLE AND FILE -- %%%%%%%%
\bibliographystyle{IEEEtranS}
\bibliography{preset.bib}
%%%%%%%%%%%%%%%%%%%%%%%%%%%%%%%%%%%%

\end{document}